\documentclass[tightenlines,eqsecnum,floats,aps,amsmath,amssymb,nofootinbib,prd,showpacs]{revtex4}
\usepackage{amsmath,amssymb,amsfonts}
\usepackage{graphicx}
\usepackage{enumerate} 
\usepackage{colordvi} 
\usepackage{bm}

\newcommand{\bfx}{\mbox{\boldmath$x$}}
\newcommand{\bfk}{\mbox{\boldmath$k$}}
\newcommand{\bfp}{\mbox{\boldmath$p$}}
\newcommand{\bfq}{\mbox{\boldmath$q$}}
\newcommand{\fnl}{f_{\rm NL}}
\newcommand{\deltag}{\delta_{\rm gal}}
\newcommand{\Pg}{P_{\rm gal}}
\newcommand{\bG}{b_{\rm\scriptscriptstyle G}}
\begin{document}
\title{Signature of Primordial Non-Gaussianity on Matter Power Spectrum}
\vfill
\author{Atsushi Taruya$^{1,2}$, Kazuya Koyama$^3$, Takahiko Matsubara$^4$}
\bigskip
\address{$^1$Research Center for the Early Universe, School of Science, 
University of Tokyo, Bunkyo-ku, Tokyo 113-0033, Japan}
\address{$^2$Institute for the Physics and Mathematics of the Universe, 
University of Tokyo, Kashiwa, Chiba 277-8568, Japan}
\address{$^3$Institute of Cosmology \& Gravitation, University of Portsmouth, 
Portsmouth, Hampshire, PO1 2EG, UK}
\address{$^4$Department of Physics, Nagoya University, Chikusa-ku, Nagoya 464-8602, Japan}
\bigskip
\date{\today}
%
\begin{abstract}
Employing the perturbative treatment of gravitational clustering, 
we discuss possible effects of primordial non-Gaussianity on 
the matter power spectrum.
As gravitational clustering develops, the 
coupling between different Fourier modes of density perturbations 
becomes important and the primordial non-Gaussianity which intrinsically 
possesses a non-trivial mode-correlation can affect the late-time 
evolution of the power spectrum. 
We quantitatively estimate the non-Gaussian effect on power spectrum 
from the perturbation theory. 
The potential impact on the cosmological parameter 
estimation using the power spectrum are investigated based on 
the Fisher-matrix formalism. 
In addition, on the basis of the local biasing prescription, 
non-Gaussian effects on the galaxy power spectrum are considered, 
showing that the scale-dependent biasing arises from a local-type
primordial non-Gaussianity. On the other hand, an equilateral-type
non-Gaussianity does not induce such scale-dependence because of
weaker mode-correlations between small and large Fourier modes.
\end{abstract}

\maketitle

\maketitle

\section{Introduction}

Recent observations of cosmic microwave background (CMB) 
anisotropies as well as density perturbations in the large-scale 
structure strongly support the basic predictions of inflationary 
scenarios, in which primordial adiabatic fluctuations 
were produced during the accelerated phase of the cosmic expansion and 
their statistical properties are approximately described by 
Gaussian statistics with a nearly scale-invariant power spectrum 
(e.g., \cite{Komatsu_WMAP5, Tegmark_etal2006}). 
Some specific inflationary models have been ruled out by 
observations, narrowing the constraints on the early stage of the 
universe. With precision measurements from future observations, 
we will detect clear signals that help us to discriminate between 
many candidate of inflationary models.

Amongst several signals accessible in the near future, 
departure from Gaussianity is an important clue to probe the generation 
mechanism for primordial perturbations as well as to 
discriminate between inflation models. 
While the simplest slow-roll inflation with single 
scalar field predicts a small departure from Gaussianity 
(e.g., \cite{ABMR2003,Maldacena2003,BKMR2004}), 
models with a non-trivial kinetic term or late-time inflaton decay called 
curvaton scenario \cite{LW2002,MT2001} can produce a large non-Gaussianity 
(e.g., \cite{SL2005,LR2005,SVW2006, Assadullahi:2007uw}). 
There are also viable models
motivated by string theory that produce large 
non-Gaussianity such as Dirac-Born-Infeld (DBI) inflation 
\cite{AST2004, Chen:2006nt, Huang:2006eh, Huang:2006eh, Arroja:2008ga, 
Langlois:2008qf, AMK2008}
and ekyprotic scenario 
\cite{KOST2001, Creminelli:2007aq, KMVW2007, Buchbinder:2007at, Lehners:2007wc}. 
Although tentative detections of primordial non-Gaussianity have been 
reported very recently (\cite{YW2008,MHLM2008}, but see also 
\cite{Komatsu_WMAP5,HMCLHM2008}) and the result  
is broadly consistent with the standard prediction of slow-roll inflation, 
these detections are at relatively low statistical significance 
and more precise measurements are necessary for 
a definite detection of the non-Gaussian signals. 
Planned CMB surveys, namely the Planck mission \cite{Planck}, 
will have much greater sensitivity and better detect non-Gaussian signals 
(e.g., \cite{SC2008}). Further, 
large-scale structure data from future spectroscopic surveys will 
uncover the mass density fluctuations up to Giga parsec scales, 
which preserve the statistical properties of primordial 
fluctuations (e.g., \cite{SSZ2004,SK2007}). 
Since these two measurements can probe different 
scales of primordial fluctuations, they are, in principle,  
complementary to each other.

In the present paper, we study the signature of primordial 
non-Gaussianity imprinted on the large-scale structure, especially 
focusing on the matter power spectrum. Naively, we expect that 
the signature of primordial non-Gaussianity basically appears in the 
statistical properties of higher-order quantities such as the bispectrum 
and trispectrum, and the power 
spectrum as a second-order statistic remains unchanged even if large 
non-Gaussianity has been produced. However, this is true only when the 
fluctuations of the mass density field are tiny and well within 
the linear regime. As the gravitational clustering develops, the coupling 
between different Fourier modes becomes important and the 
scale-dependent non-linear growth appears  
due to the mode-coupling 
effect. In the weakly non-linear regime, the strength of 
this mode-coupling sensitively depends on the initial condition. Since 
the non-Gaussian density field intrinsically possesses non-trivial 
mode-correlations, the evolution of the power spectrum may be altered by the 
primordial non-Gaussianity, at least in the weakly non-linear regime.

Here, employing the perturbative calculations, 
we study the effect of primordial non-Gaussianity on the matter 
power spectrum. 
The effect of non-Gaussianity on the matter power spectrum 
has been previously investigated by Ref.\cite{SSZ2004} using perturbation 
theory, showing that the magnitude of this effect is roughly at 
the few percent level in the weakly non-linear regime. 
In this paper, adopting the two representative models of primordial 
non-Gaussianity described in Sec.~\ref{sec:nonGaussian_model}, 
we quantitatively estimate  
the non-Gaussian effects on the power spectrum and elucidate their shape 
dependence (Sec.~\ref{sec:non_GaussianPT}). 
In particular, we 
carefully examine the influence of primordial non-Gaussianity on 
cosmological parameter estimation, paying particular attention to 
the primordial spectral indices and the characteristic scale of the 
baryon acoustic oscillations as a cosmic standard ruler 
(Sec.~\ref{sec:Fisher}). 
Also, we discuss the non-Gaussian effect on the galaxy biasing 
(Sec.~\ref{sec:galaxy_biasing}). This issue has been recently studied by 
several authors in the context of the halo biasing prescription 
\cite{DDHS2008,MV2008,SHSHP2008,McDonald2008,AT2008}, 
and local biasing prescription \cite{McDonald2008}. 
Here, we adopt the local biasing scheme and 
consider the scale-dependent galaxy biasing arising from the 
primordial non-Gaussianity.

Throughout the paper,  
we assume a flat Lambda cold dark matter (CDM) model and the 
fiducial model parameters are chosen based on the five-year WMAP results 
(WMAP+BAO+SN ML, see Ref.\cite{Komatsu_WMAP5}): $\Omega_{\rm b}h^2=0.02263$, 
$\Omega_{\rm c}h^2=0.1136$, 
$\Omega_{\rm K}=0\,\, (\Omega_{\Lambda}=0.724)$, $n_s=0.961$, 
$\tau=0.080$, $\Delta_{\mathcal R}(k=0.002\mathrm{Mpc}^{-1})
=2.42\times10^{-9}$, $h=0.703$.

\section{Primordial non-Gaussianity}
\label{sec:nonGaussian_model}

According to the commonly accepted mechanisms to 
generate the primordial fluctuations, 
non-Gaussianity would be imprinted on the 
primordial curvature perturbation produced during inflation. 
The curvature perturbation can also be generated after 
inflation by a late-time decay of light fields, and  
this would produce a large non-Gaussianity in the 
curvature perturbation. Further, as alternative scenarios to inflation 
such as ekpyrotic models, nearly scale-invariant 
curvature perturbation with large non-Gaussianity could be generated 
during the collapsing phase of the Universe.

Let us denote the primordial curvature perturbation 
on the uniform density hypersurface and 
the Bardeen's curvature potential at super-horizon scales 
by $\zeta_p$ and $\Phi_{H,p}$, respectively. 
The primordial density perturbation defined at an early time of 
the matter dominated epoch, $\delta_0$, is related to these variables 
through (e.g., \cite{HKM2006,LMSV2007,SSZ2004,SK2007})
\begin{equation}
\delta_0(\bfk)=M_\zeta(k)\zeta_p(\bfk) 
= \frac{5}{3}M_\zeta(k)\Phi_{H,p}(k), 
\label{eq:delta0_Phi_p}
\end{equation}
with the function $M_\zeta$ being
\begin{equation}
M_\zeta(k)= \frac{2}{5}\frac{k^2\,T(k)}{\Omega_{\rm m,0}H_0^2}, 
\label{eq:M_zeta}
\end{equation}
where the function $T(k)$ is the transfer function of matter fluctuations  
normalized to unity at $k\to0$. 
Note that the curvature potential given in the matter dominated epoch, 
$\Phi_{H,p}$, is related to the curvature perturbation
$\zeta(\bfk)=(5/3)\,\Phi_{H,p}(\bfk)$.

Using the relation (\ref{eq:delta0_Phi_p}), $n$-point statistics of 
the primordial density field are all expressed in terms of 
respective correlator of the curvature potential or the curvature
perturbation. For the power spectra, we have
\begin{eqnarray}
P_0(k)=M_\zeta^2(k)\,P_\zeta(k), 
\end{eqnarray}
with quantities $P_0$ and $P_\zeta$ defined by  
\begin{equation}
\langle\delta_0(\bfk)\delta_0(\bfk')\rangle=
(2\pi)^3\delta_{\rm D}(\bfk+\bfk')\,P_0(k),
\quad
\langle\zeta_p(\bfk)\zeta_p(\bfk')\rangle=
(2\pi)^3\delta_{\rm D}(\bfk+\bfk')\,P_{\zeta}(k).
\label{eq:P_primordial(k)}
\end{equation}
The bispectrum as the first non-trivial quantity arising from the 
non-Gaussianity becomes 
\begin{eqnarray}
B_0(\bfk_1,\bfk_2,\bfk_3)=M_\zeta(k_1)M_\zeta(k_2)M_\zeta(k_3)\,\,
B_\zeta(\bfk_1,\bfk_2,\bfk_3), 
\end{eqnarray}
where we define 
\begin{eqnarray}
\Bigl\langle\delta_0(\bfk_1)\delta_0(\bfk_2)\delta_0(\bfk_3)\Bigr\rangle 
&=& 
(2\pi)^3\,\delta_{\rm D}(\bfk_1+\bfk_2+\bfk_3)\,B_0(\bfk_1,\bfk_2,\bfk_3),
\nonumber\\
\Bigl\langle\zeta_p(\bfk_1)\zeta_p(\bfk_2)\zeta_p(\bfk_3)\Bigr\rangle 
&=& 
(2\pi)^3\,\delta_{\rm D}(\bfk_1+\bfk_2+\bfk_3)\,B_\zeta(\bfk_1,\bfk_2,\bfk_3).
\end{eqnarray}
Details of the functional form of $B_\zeta$ depend on the generation 
mechanism of primordial non-Gaussianity.
In the following, we will consider two representative 
models of non-Gaussianity and give the explicit expressions for 
the primordial bispectrum $B_\zeta$.

\subsection{Local model}

In the local model of primordial non-Gaussianity, 
the primordial fluctuation $\Phi_{H,p}$
is characterized by the Taylor expansion of the Gaussian field. 
Denoting the Gaussian field by $\varphi$, 
it is conventionally characterized as: 
\begin{equation}
\Phi_{H,p}(\bfx)=\varphi(\bfx)+ 
\fnl\left\{\varphi^2(\bfx)-\langle\varphi^2\rangle\right\}+ 
\cdots.
\end{equation}
The Fourier counterpart of the above equation is given by  
\begin{equation}
\Phi_{H,p}(\bfk)=\varphi(\bfk)+ 
\fnl\,\int\frac{d^3\bfq}{(2\pi)^3}
\left\{\varphi(\bfq)\varphi(\bfk-\bfq)-
\langle\varphi(\bfq)\varphi(\bfk-\bfq)\rangle\right\} 
+\cdots.
\label{eq:expand_Phi_p}
\end{equation}
Alternatively, we can expand $\zeta_p$ as: 
\begin{equation}
\zeta_{p}(\bfk)= \zeta_{\rm G}(\bfk) + 
\frac{3}{5}\fnl\,\int\frac{d^3\bfq}{(2\pi)^3}
\left\{\zeta_{\rm G}(\bfq)\zeta_{G}(\bfk-\bfq)-
\langle\zeta_{\rm G}(\bfq)\zeta_{\rm G}(\bfk-\bfq)\rangle\right\} 
+\cdots,
\end{equation}
where $\zeta_{\rm G}$ is Gaussian field and we have the relation 
$\zeta_{\rm G}=(5/3)\varphi$. For the cases of interest here, 
deviation from Gaussianity is generally small and perturbative 
evaluation of the primordial bispectrum is valid. The leading-order 
result becomes
\begin{eqnarray}
B_\zeta(\bfk_1,\bfk_2,\bfk_3)\,\simeq\,\frac{6}{5}\fnl\,
\Bigl[ P_\zeta(k_1)P_\zeta(k_2)+P_\zeta(k_2)P_\zeta(k_3)+
P_\zeta(k_3)P_\zeta(k_1)\Bigr].
\label{eq:bispectrum_local}
\end{eqnarray}
In the slow-roll inflation models, $\fnl$ is suppressed by 
the slow-roll parameters and thus it is expected to be very small.  
However, in curvaton models or ekpyrotic models, it is possible to 
generate a large non-Gaussianity of the local model with 
$\fnl \sim \mathcal{O}(10^2)$.

\subsection{Equilateral model}

As another type of bispectrum configuration, we consider 
an equilateral model. The DBI inflation is a typical 
example where the bispectrum has a maximum amplitude for 
equilateral configurations. This is in contrast with the 
local model where the bispectrum has the largest amplitude
for squeezed configurations \cite{BCM2004}. The bispectrum in the equilateral 
model is approximated by 
\cite{CSZT2007}\footnote{The expression given here is 
slightly generalized in a sense that we do not necessarily assume 
a power-law form of the power spectrum, $P_\zeta(k)$.}
\begin{eqnarray}
B_\zeta(\bfk_1,\bfk_2,\bfk_3)&=&\frac{18}{5}\,\,\fnl^{\rm eq}
\left(\frac{\bar{k}}{k_{\rm CMB}}\right)^{-2\kappa}
\,\,\Bigl[ -P_\zeta(k_1)P_\zeta(k_2)-P_\zeta(k_2)P_\zeta(k_3)
-P_\zeta(k_3)P_\zeta(k_1)
\nonumber\\
&& \quad
\Bigl.-2\,\{P_\zeta(k_1)P_\zeta(k_2)P_\zeta(k_3)\}^{2/3}
+ \left\{ \Bigl(P_\zeta(k_1)\{P_\zeta(k_2)\}^2\{P_\zeta(k_3)\}^3\Bigr)^{1/3}
+(\mathrm{5\,\,perm.})\right\}\,\,\Bigr],
\label{eq:bispectrum_eq}
\end{eqnarray}
with $\bar{k}=(k_1+k_2+k_3)/3$. 
Here, we introduce the slope index $\kappa$ in order to 
characterize the scale-dependence of the non-Gaussianity, which 
generically appears in the DBI-type inflation models. 
we set the characteristic scale to $k_{\rm CMB}=0.04\,\mbox{Mpc}^{-1}$.
Note that the bispectrum in equation (\ref{eq:bispectrum_eq}) is 
normalized in such a 
way that, for equilateral configurations ($k_1=k_2=k_3=k$), 
it coincides with the local form given in equation 
(\ref{eq:bispectrum_local}).

\section{Non-linear Evolution of Matter Power Spectrum 
from Perturbation Theory}
\label{sec:non_GaussianPT}

\subsection{Perturbation  Theory}

Even if the primordial fluctuations are well described by 
linear theory, the non-linearity of the gravitational dynamics 
eventually dominates and we must correctly take into account the 
non-linear growth of matter fluctuations, which significantly modifies the 
matter power spectrum. For the scales of interest here, especially the 
accessible scales of future galaxy redshift surveys, 
the non-linear evolution is rather moderate and 
perturbative treatment is valid. 
The mass density field is described by 
\begin{equation}
\delta(\bfk;z) = \delta^{(1)}(\bfk;z)+\delta^{(2)}(\bfk;z)+
\delta^{(3)}(\bfk;z)+\cdots.
\end{equation}
In the above, the function $\delta^{(1)}(\bfk;z)$ represents the 
linear fluctuation and it is given by 
$\delta^{(1)}(\bfk;z)=D(z)\delta_0(\bfk)$, where $D(z)$ is the linear 
growth rate. Neglecting the decaying mode of linear perturbation, 
the solutions for higher-order quantities are formally expressed as 
\cite{BCFS2002}
\begin{eqnarray}
\delta^{(n)}(\bfk;z)=[D(z)]^n \,\int\frac{d^3k_1\cdots d^3k_n}
{(2\pi)^{3(n-1)}}\,\delta_{\rm D}(\bfk-\bfk_1-\cdots-\bfk_n)\,
F^{(n)}_{\rm sym}(\bfk_1,\cdots,\bfk_n)\delta_0(\bfk_1)\cdots\delta_0(\bfk_n),
\label{eq:PT_sol}
\end{eqnarray}
with $F_{\rm sym}^{(n)}$ being the symmetrized kernel for the $n$-th order 
solution. Then, the power spectrum of the density field, defined as 
\begin{equation}
\langle\delta(\bfk;z)\delta(\bfk';z)\rangle=
(2\pi)^3\delta_{\rm D}(\bfk+\bfk')\,\,P(k;z),
\end{equation}
can be calculated by substituting Eq.(\ref{eq:PT_sol}) into the above 
expression. The resultant expression is summarized as  
\begin{equation}
P(k;z) = D^2(z)\,P_0(k) + P^{(12)}(k;z) + 
\left[\,P^{(22)}(k;z) + P^{(13)}(k;z)\, \right]\,+\cdots,
\end{equation}
where the terms $P^{(12)}$, $P^{(22)}$ and $P^{(13)}$ are the so-called 
one-loop power spectra given by 
\begin{eqnarray}
P^{(12)}(k;z)&=&2\,D^3(z)\,\int \frac{d^3q}{(2\pi)^3}\,
F^{(2)}_{\rm sym}(\bfq,\bfk-\bfq)\,B_0(-\bfk,\bfq,\bfk-\bfq),
\label{eq:P12}
\\
P^{(22)}(k;z)&=&2\,D^4(z)\,\int \frac{d^3q}{(2\pi)^3}\,
\{\,F^{(2)}_{\rm sym}(\bfq,\bfk-\bfq)\,\}^2\,P_0(q)P_0(|\bfk-\bfq|)
\nonumber\\
&&+ D^4(z)\,\int \frac{d^3p d^3q}{(2\pi)^6}\,
F^{(2)}_{\rm sym}(\bfp,\bfk-\bfp)F^{(2)}_{\rm sym}(\bfq,-\bfk-\bfq)\,
T_0(\bfp,\bfk-\bfp,\bfq,-\bfk-\bfq),
\label{eq:P22}
\\
P^{(13)}(k;z)&=&6\,D^4(z)\,\int \frac{d^3q}{(2\pi)^3}\,
F^{(3)}_{\rm sym}(\bfk,\bfq,-\bfq)\,P_0(k)P_0(q)
\nonumber\\
&&+2\,D^4(z)\,\int\frac{d^3p d^3q}{(2\pi)^6} \,
F^{(3)}_{\rm sym}(\bfp,\bfq,\bfk-\bfp-\bfq)\,
T_0(-\bfk,\bfp,\bfq,\bfk-\bfp-\bfq). 
\label{eq:P13}
\end{eqnarray}
These are the most general expressions for the one-loop power spectra 
in the presence of primordial non-Gaussianity. 
The quantity $P^{(12)}$ is the first 
non-trivial correction associated with the primordial density bispectrum 
$B_0(\bfk_1,\bfk_2,\bfk_3)$. 
While the quantities $P^{(22)}$ and 
$P^{(13)}$ represent the first leading-order corrections in the case of 
Gaussian initial conditions, additional contributions  
originating from the primordial density trispectrum, 
$T_0(\bfk_1,\bfk_3,\bfk_3,\bfk_4)$, also arise. 
Thus, in general, we need the explicit functional form of the primordial 
trispectrum as well as primordial bispectrum in order to compute the 
one-loop power spectra. 
However, in most general cases, the primordial trispectrum is 
of the order of $M_\zeta^4P_0^3$, which is negligibly smaller than 
the bispectrum, $B_0\sim M_\zeta^3P_0^2$. 
Since the terms including $P_0^3$ may be regarded as the two-loop 
order, we can safely neglect the trispectrum contribution. 
With this treatment, the terms $P^{(22)}$ and $P^{(13)}$ reduce to nothing but
the {\it standard} one-loop spectra, and they are explicitly given by 
(e.g., \cite{MSS1992,JB1994,SF1996,Nishimichi2007})
\begin{eqnarray}
P^{(22)}(k;z)&=&D^4(z)\,\frac{k^3}{(2\pi)^2}\,\int_0^{\infty} dx\, 
x^2 P_0(kx) \int_{-1}^{+1} d\mu\,P_0 
\left(k\sqrt{1+x^2-2\mu\,x}\right)\, \frac{1}{2}
\left[\frac{3x+7\mu-10\mu^2x}{7x(1+x^2-2\mu x)}\right]^2,
\label{eq:P_ab^22(k)}
\\
P^{(13)}(k;z)&=&D^4(z)\,\frac{k^3}{(2\pi)^2}\,\,P_0(k)\,
\int_0^{\infty} dx\, x^2 P_0(kx) 
\nonumber\\
&&\quad\quad\times\,\,\frac{1}{252\,x^2}
\left[\frac{12}{x^2}-158+100x^2-42x^4+
\frac{3}{x^3}\,(x^2-1)^3(7x^2+2)\,\ln\left|\frac{x+1}{x-1}\right|
\right].
\label{eq:P_ab^13(k)}
\end{eqnarray}
Also, the non-trivial contribution from the primordial bispectrum is 
expressed in terms of the bispectrum of the curvature perturbation as  
\begin{eqnarray}
P^{(12)}(k;z)&=& D^3(z)\,\frac{k^3}{(2\pi)^2} 
\int_0^{\infty}dx x^2 \int_{-1}^{+1}d\mu \,\,
\frac{3x+7\mu-10\mu^2\,x}{7x(1+x^2-2\mu\,x)}\,
\nonumber\\
&&\quad\quad\times\,\,M_\zeta(k)M_\zeta(kx)M_\zeta(k\sqrt{1+x^2-2\mu x})\,\,
B_\zeta(k,kx,k\sqrt{1+x^2-2\mu x}).
\label{eq:P_ab^12(k)}
\end{eqnarray}

\begin{figure}[t]
\begin{center}
\includegraphics[width=8.5cm,angle=0]{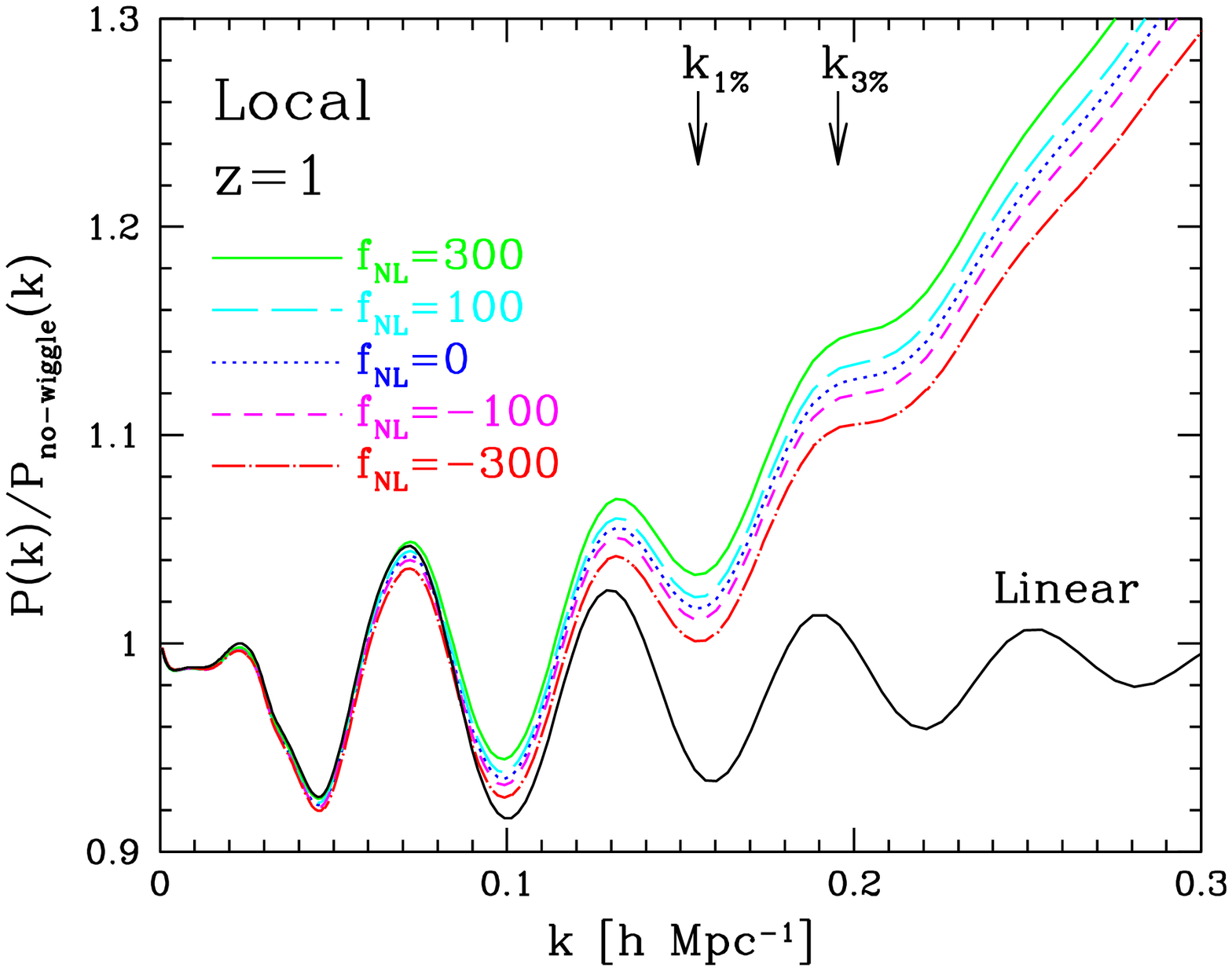}
\hspace*{0cm}
\includegraphics[width=8.5cm,angle=0]{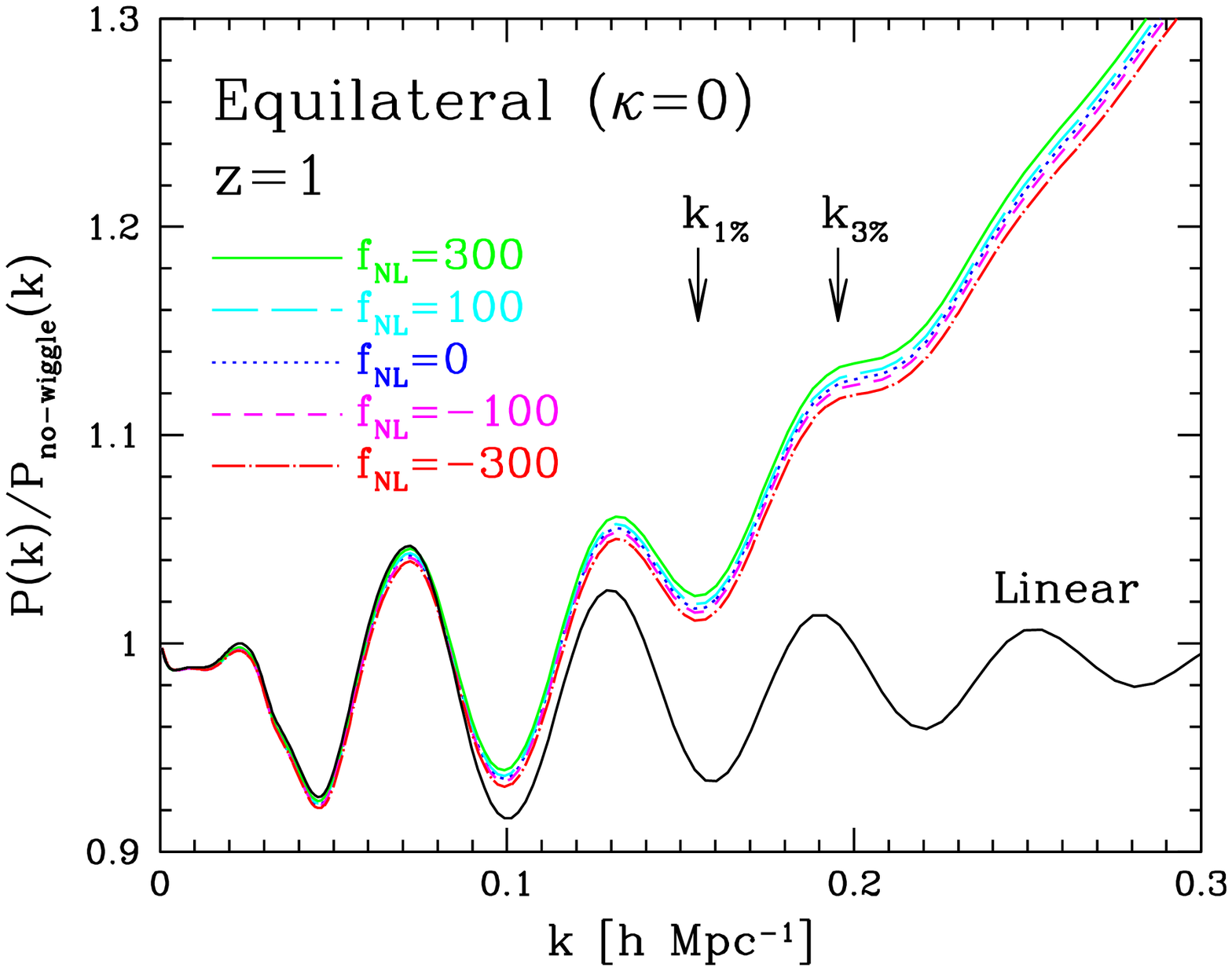}
\end{center}
\caption{Ratio of the power spectrum to the smoothed reference spectrum for 
  the local model (left) and the equilateral model 
  with $\kappa=0$ (right), for $z=1$. 
  The smooth spectra are obtained from the no-wiggle 
  approximation of the linear transfer function according to 
  Ref.~\cite{EH1998}. 
  The lines from top to bottom respectively indicate $\fnl=300$ (solid), 
  $\fnl=100$   (long-dashed), $\fnl=0$ (dotted), $\fnl=-100$ (short-dashed) 
  and $\fnl=-300$ (dot-dashed). 
\label{fig:ratio_nowiggle}}
\end{figure}

\subsection{Results}

\begin{figure}[t]
\begin{center}
\includegraphics[width=10.5cm,angle=0]{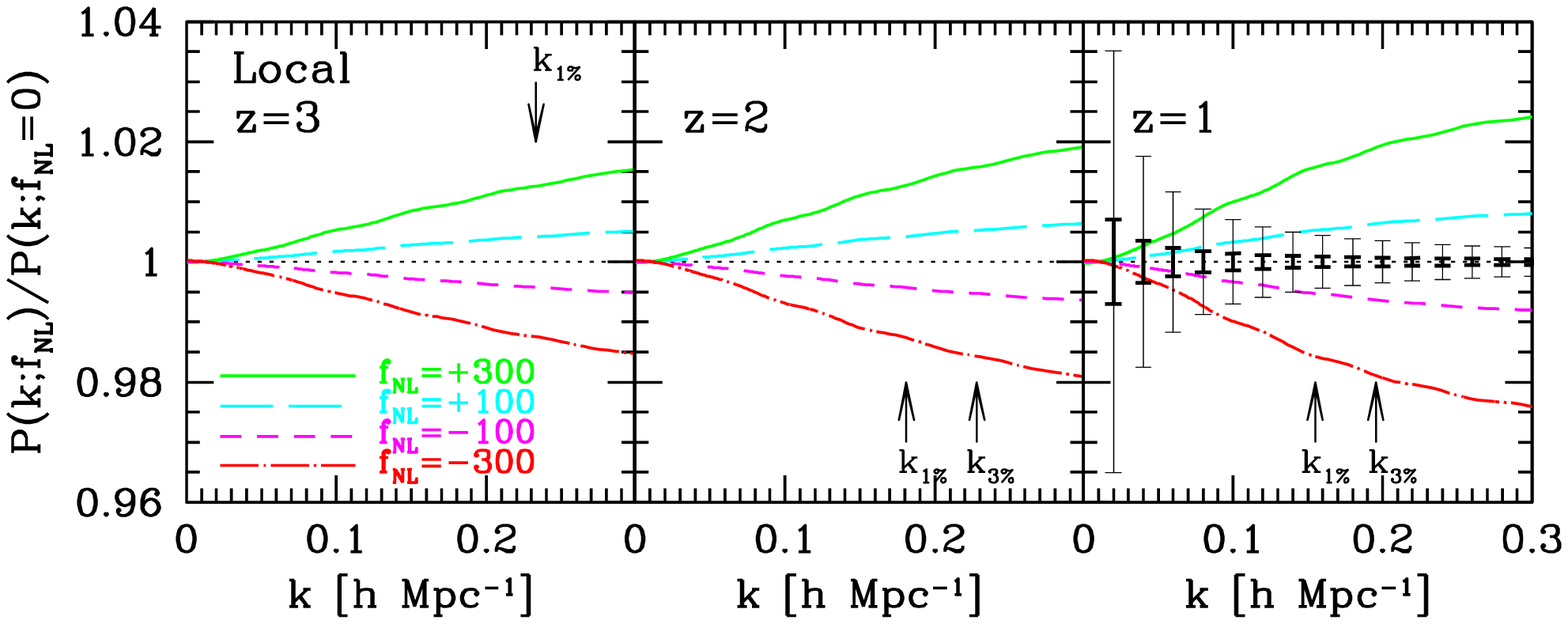}

\vspace*{0.1cm}

\includegraphics[width=10.5cm,angle=0]{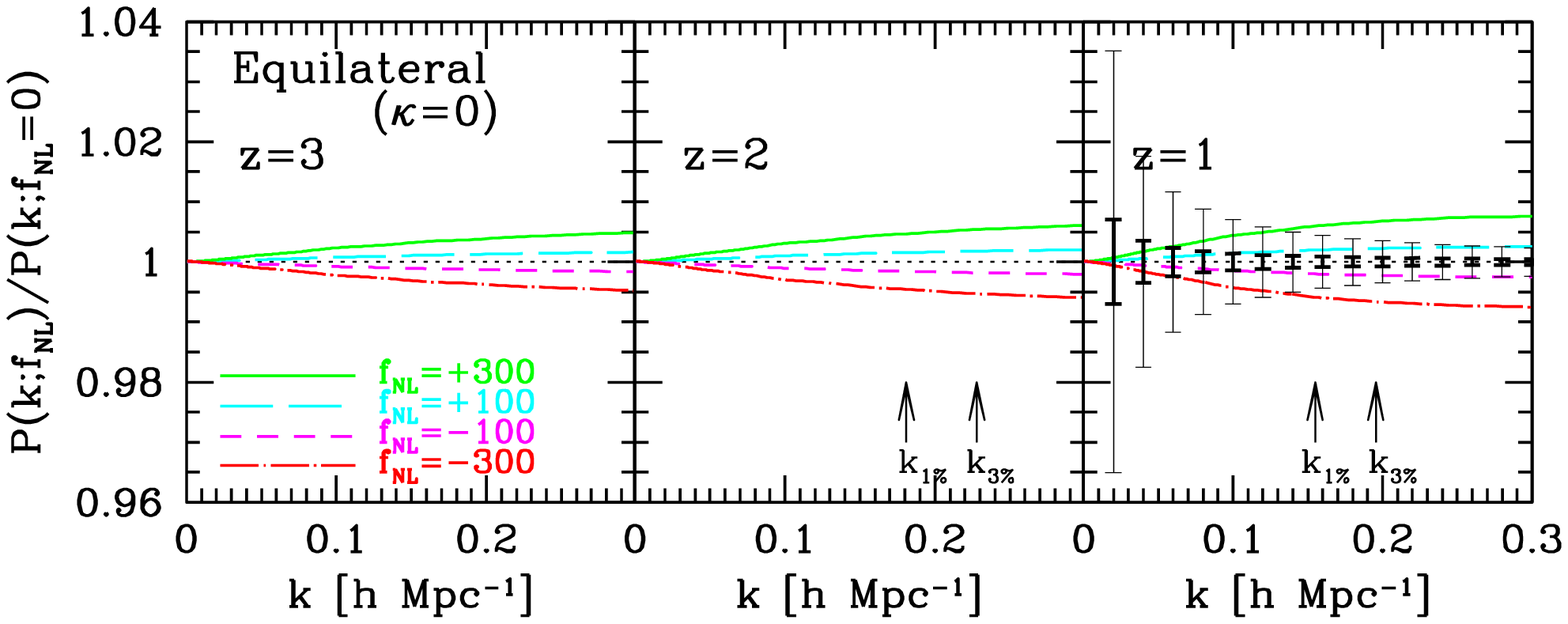}

\vspace*{0.1cm}

\includegraphics[width=10.5cm,angle=0]{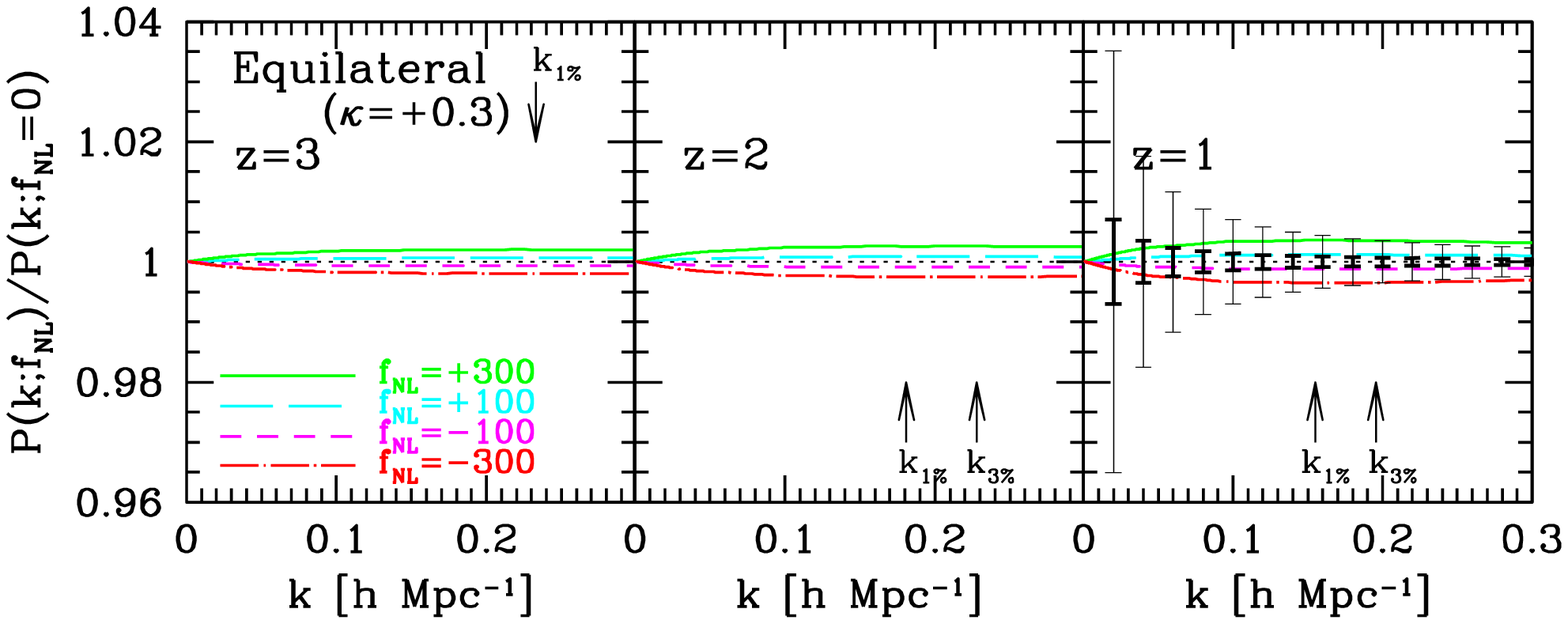}

\vspace*{0.1cm}

\includegraphics[width=10.5cm,angle=0]{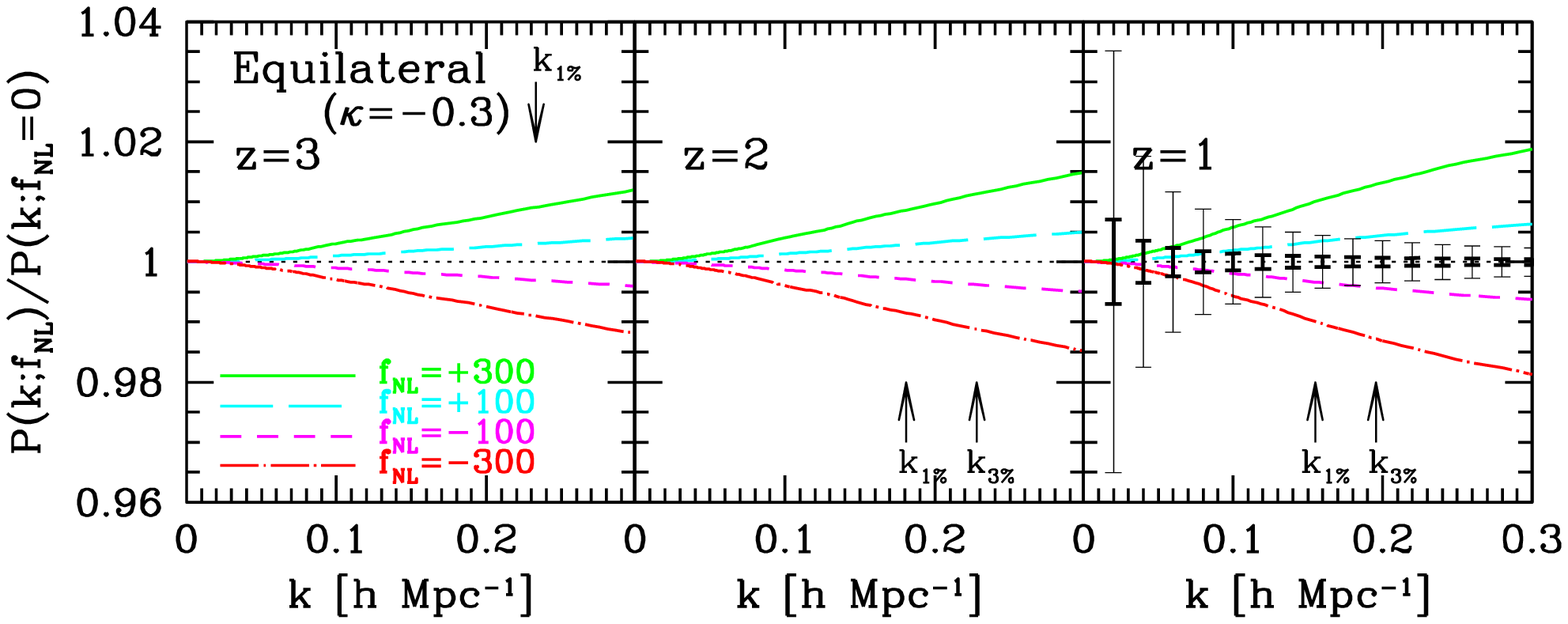}
\end{center}

\vspace*{-0.4cm}

\caption{Ratios of power spectrum, $P(k;\fnl\ne0)/P(k;\fnl=0)$, 
for $z=3$ (left), $2$ (middle) and $1$ (right). 
From top to bottom panels, the results for the local, equilateral with 
$\kappa=0, ~~0.3$, and $-0.3$ are shown, respectively. 
In each panel, we plot the cases with non-Gaussian parameter 
$\fnl=+300$ (solid), $+100$ (long-dashed), 
$-100$ (short-dashed) and $-300$ (dot-dashed). 
The vertical arrows labeled by $k_{1\%}$ and $k_{3\%}$ indicate 
the maximum wave number below which the perturbation theory predictions 
are reliable with a 
precision of $1\%$ and $3\%$ level, respectively, according to 
the criteria (\ref{eq:criteria}) \cite{Nishimichi2008}.  
As references, the error bars limited by 
the cosmic variance are plotted in the right panel, assuming the survey 
volume of $V_s=4\, (h^{-1}$Gpc$)^3$ (thin solid) and 
$V_s=10^2\, (h^{-1}$Gpc$)^3$ (thick solid). 
\label{fig:ratio_pk}}
\end{figure}

Based on the expressions given above, 
we now calculate the matter power spectrum,   
focusing on the scales relevant for future galaxy surveys. 
Figure \ref{fig:ratio_nowiggle} plots the ratios of power spectra 
to the smooth reference spectrum, 
$P(k)/P_{\rm no\mbox{-}wiggle}(k)$, 
given at $z=1$. The reference spectrum $P_{\rm no\mbox{-}wiggle}(k)$ is 
the linear power spectra computed from 
the no-wiggle approximation of the transfer function given 
by Ref.~\cite{EH1998}. The left panel shows the results for 
the local model, and the right panel plots the results for 
the equilateral model with index $\kappa=0$. 
The overall behavior at higher $k$ is basically dominated by the 
corrections coming from the standard 1-loop power spectra,  
$P^{(22)}$ and $P^{(13)}$, but small variation of the power spectrum 
amplitudes manifests at higher $k$. This is 
the contribution of primordial non-Gaussianity, $P^{(12)}$, 
originating from the primordial bispectrum. 
Depending on the sign of $\fnl$, the primordial 
non-Gaussianity enhances or suppresses the non-linear growth of the power 
spectrum and the deviation from the Gaussian result $(\fnl=0)$ becomes 
more significant as $|\fnl|$ increases. 
Although the effect of primordial non-Gaussianity seems rather mild 
within the current constrained values of 
$-9\lesssim\fnl^{\rm local}\lesssim111$ and 
$-151\lesssim\fnl^{\rm eq}\lesssim253$ (95\%C.L.)\cite{Komatsu_WMAP5}, 
we do not immediately exclude the possibility of significant impact on  
the cosmological parameter estimation. We will discuss this issue 
in the next section.

Next we will focus on the shape of the power spectrum. In 
Figure \ref{fig:ratio_pk}, the ratios of the 
power spectra to those in the Gaussian case ($\fnl=0$), i.e., 
$P(k;\fnl\ne0)/P(k;\fnl=0)$, are plotted for 
various redshifts and values of $\fnl$. Note that in computing the 
ratio, both the numerator and denominator are calculated from perturbation 
theory. 
In Figures \ref{fig:ratio_nowiggle} and \ref{fig:ratio_pk}, 
the vertical arrows labeled by 
$k_{1\%}$ and $k_{3\%}$ indicate the maximum wavenumber below which the 
precision level of perturbation theory prediction 
is expected to be better than $1\%$ and $3\%$, respectively. 
According to recent numerical experiments, the maximum wavenumber 
is empirically determined through \cite{Nishimichi2008}
\begin{equation}
k^2\int_0^{k} \frac{dq}{6\pi^2} \,D^2(z)P_0(q)=C,  
\label{eq:criteria}
\end{equation}
where $C=0.18~(0.3)$ for $k_{1\%}~(k_{3\%})$. 
Solving the above equation with respect to $k$, we obtain 
$k_{1\%}$ and $k_{3\%}$ as function of redshift. 
Note that the maximum wave numbers given above have been derived 
by comparison between N-body simulations and theoretical predictions, 
and the resultant convergence range is narrower than 
those previously proposed (e.g., \cite{JK2006,SK2007}). In this sense, 
Eq.(\ref{eq:criteria}) 
may be regarded as a conservative criterion.

Bearing the limitation of perturbation theory in mind, we see that 
the deviation of the power spectrum from the Gaussian case also depends on the 
redshift and scales as well as the model of primordial non-Gaussianity.  
Monotonic change of the power spectrum amplitude 
is broadly consistent with recent N-body simulations in the case of the 
local model \cite{GBDMM2008}, 
although the fractional change is 
typically $\lesssim 2-3\%$, as expected from Figure \ref{fig:ratio_nowiggle}. 
Note, however, that the level of this changes roughly corresponds to 
the sensitivity achievable with future redshift surveys. 
As references, we plot the error bars in each right panel, 
which represent the expected $1-\sigma$ errors, $\Delta P(k)$,  
limited by the cosmic-variance, i.e.,  
$\Delta P(k)/P(k)=(2/N_k)^{1/2}$ with $N_k$ being the number of 
Fourier modes within a given bin at $k$. Here, we specifically 
consider the two representative cases of $V_{\rm s}=4\,(h^{-1}$Gpc$)^3$ 
(thin) and $V_{\rm s}=10^2\,(h^{-1}$Gpc$)^3$ (thick), 
corresponding to the future surveys dedicated for the measurement of 
baryon acoustic oscillations (BAOs) from the ground and 
space, respectively. 
Naively comparing the error bars to the amplitudes of the 
non-Gaussian effects, ground-based BAO surveys with a typical volume of 
$V\sim$few $h^{-3}$Gpc$^3$ will find it difficult to detect the signature of 
primordial non-Gaussianity only from the power spectrum. However, 
idealistic surveys with huge volumes seem to show the potential 
for a definite detection of non-Gaussian effects even with 
the currently constrained values of $\fnl$.


\section{Impact on Cosmological Parameter Estimation}
\label{sec:Fisher}

\subsection{Fisher-matrix analysis}

Apart from primordial non-Gaussianity, there are several parameters 
that affect the shape of the power spectrum. Among these, the primordial 
spectral index $n_s$ and running index $\alpha\equiv dn_s/d\ln k$ 
monotonically change the power spectrum, which resemble 
the effect of primordial non-Gaussianity at some wavenumbers.  
A natural question arises whether the primordial non-Gaussianity is 
indeed detectable or not considering the degeneracies 
and how the non-negligible effect of 
non-Gaussianity adversely affects the estimations of primordial spectral 
index and the running of the index. 
On the other hand, for 
the scales accessible to the future surveys, a measurement of the 
characteristic scale of BAOs is an important clue to probe the 
late-time acceleration 
of the universe e.g., \cite{HH2003,SE2003,Matsubara2004}), 
and a percent-level determination of the acoustic scale is 
required for the determination of the dark energy equation of state 
\cite{A2006,P2006}. In this respect, even the small effect 
of primordial non-Gaussianity may affect the determination of the 
BAO scale.

Here, we apply the Fisher-matrix method to address 
these issues and explore the potential impact on 
cosmological parameter estimation as well as the determination of 
the BAO scale. For this purpose, we consider a rather simplified setup, 
namely that our observable is the real-space power spectrum 
free from the redshift-space distortion. Under the assumption of 
linear galaxy biasing, the observed power spectrum may be written as 
(e.g., \cite{JK2006}): 
\begin{equation}
P_{\rm obs}(k;z)= \left(\frac{D_V(z)}{D_{V,{\rm true}}(z)}\right)^3 \,
b_1^2\,P_{\rm mass}\left(\frac{D_{V,{\rm true}}(z)}{D_V(z)}\,k;\,z\right),
\label{eq:pk_obs}
\end{equation}
where the power spectrum $P_{\rm mass}$ is computed from perturbation 
theory. Here, $b_1$ is the linear biasing parameter, and $D_V(z)$ represents 
the cosmological distance averaged over three-dimensional space, 
$D_V(z)\equiv[D_A^2(z)/H(z)]^{1/3}$. The subscript ``true'' denotes the  
quantities estimated from the true cosmological model.

Following Refs.\cite{TTH1997,Tegmark1997}, the Fisher matrix for the 
galaxy power spectrum is given by 
\begin{eqnarray}
F_{ij}=\frac{V_s}{(2\pi)^2}\int_{k_{\rm min}}^{k_{\rm max}}dk\,k^2\,
\frac{\partial \ln P(k;z)}{\partial \theta_i}
\frac{\partial \ln P(k;z)}{\partial \theta_j}
\left\{\frac{n_{\rm gal}\,P(k;z)}{n_{\rm gal}\,P(k;z)+1}\right\}^2,
\label{eq:fisher}
\end{eqnarray}
where $\theta_i$ represents one from a set of parameters. The quantity 
$V_s$ is the survey volume and $n_{\rm gal}$ is the number density 
of galaxies. The range of integration $[k_{\rm min}, k_{\rm max}]$ 
should be determined through the survey properties, and in particular 
the minimum wave number is limited to $2\pi/V_s^{1/3}$. 
Note that the choice of $k_{\rm min}$ may be crucial for 
the power spectrum measurement on large-scales $k\lesssim0.01h$Mpc$^{-1}$, 
where the scale-dependent effect of galaxy biasing becomes prominent 
through the non-trivial mode-coupling from primordial non-Gaussianity 
\cite{DDHS2008,MV2008,SHSHP2008,McDonald2008,AT2008}. 
Here, we fix $k_{\rm min}=0.01h$Mpc$^{-1}$ and allow $k_{\rm max}$ to vary. 
The effect of scale-dependent biasing will be discussed in next section.

In the Fisher-matrix analysis presented below, we include the five 
parameters given by 
$\theta_i=\{n_s,\,\,\alpha,\,\,D_V/D_{V,{\rm true}},\,\,\fnl,\,\, b_1\}$. 
The fiducial values of these parameters 
are set to $n_s=0.961$, $\alpha=0$, $D_V/D_{V,{\rm true}}=1$, and $\fnl=0$. 
We then consider a space-based BAO 
survey as an idealistically gigantic galaxy survey along the lines of 
SPACE and ADEPT 
(\cite{DETF,SPACE,ADEPT}, see also Ref.~\cite{SK2007}). 
We adopt the survey parameters of a space-based BAO experiment as 
follows: $z=1.5$, $V_s=100$\,($h^{-1}$Gpc)$^3$, 
$b_1=3.25$, and $n_{\rm gal}=3.25\times10^{-4}$($h^{-1}$Mpc)$^{-3}$ 
(e.g., \cite{SK2007}). 
Below, we mainly focus on the local model of primordial non-Gaussianity. 
The results for the equilateral model are qualitatively the same as in 
the local model and are briefly summarized in Table \ref{tab:fisher_results}.

\subsection{Results}

\begin{figure}[t]
\begin{center}
\includegraphics[width=5.7cm,angle=0]{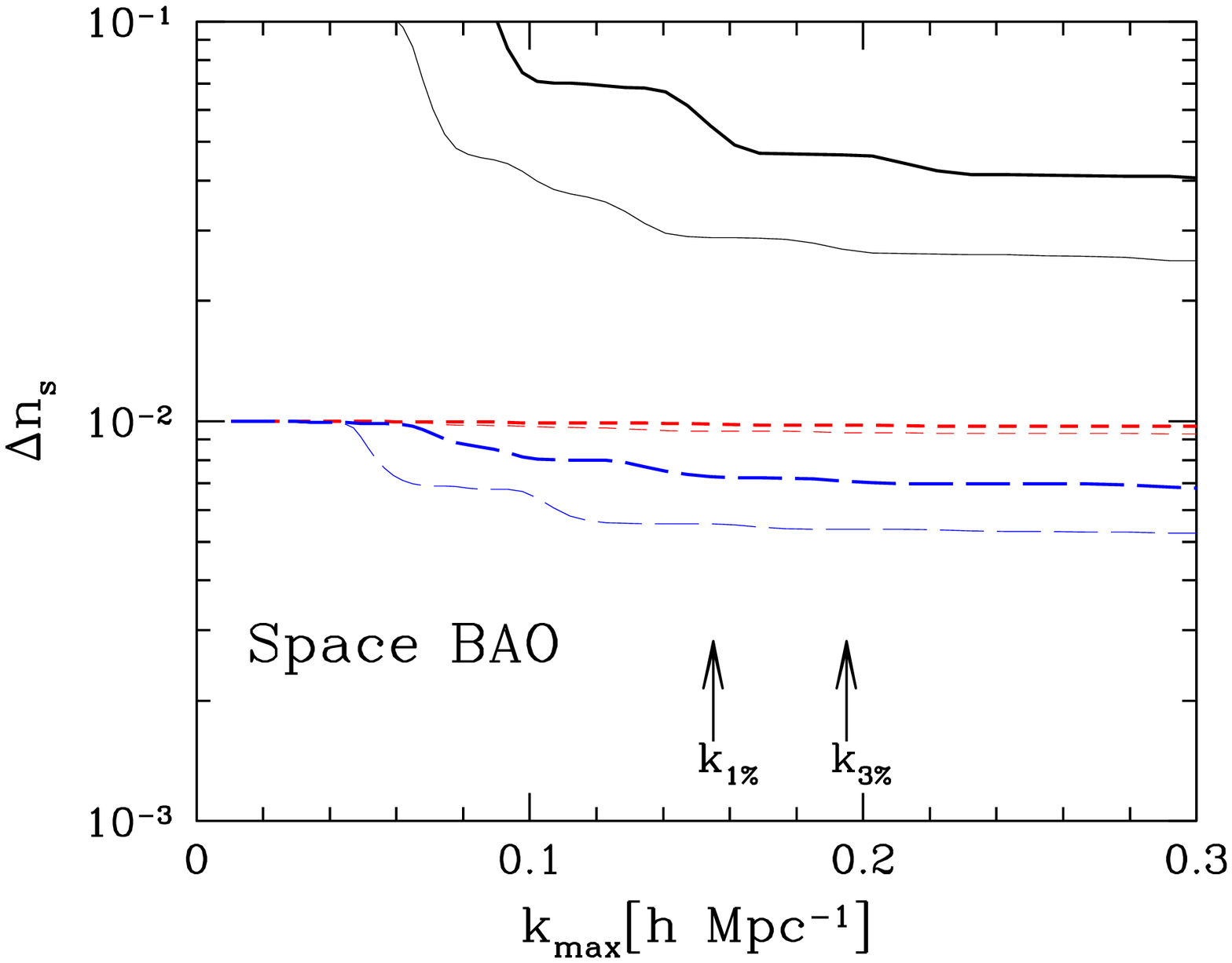}
\hspace*{0.cm}
\includegraphics[width=5.7cm,angle=0]{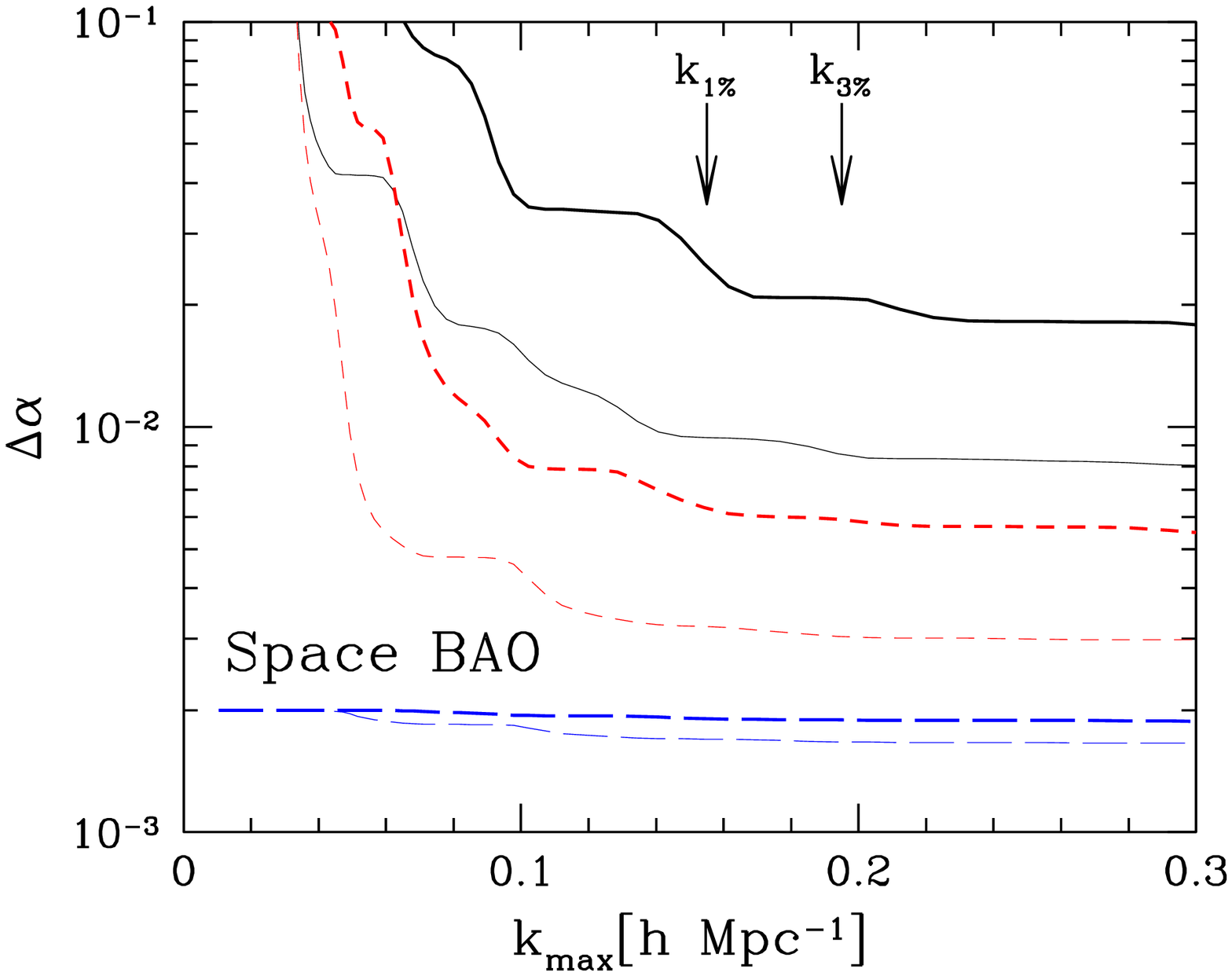}
\hspace*{0.cm}
\includegraphics[width=5.7cm,angle=0]{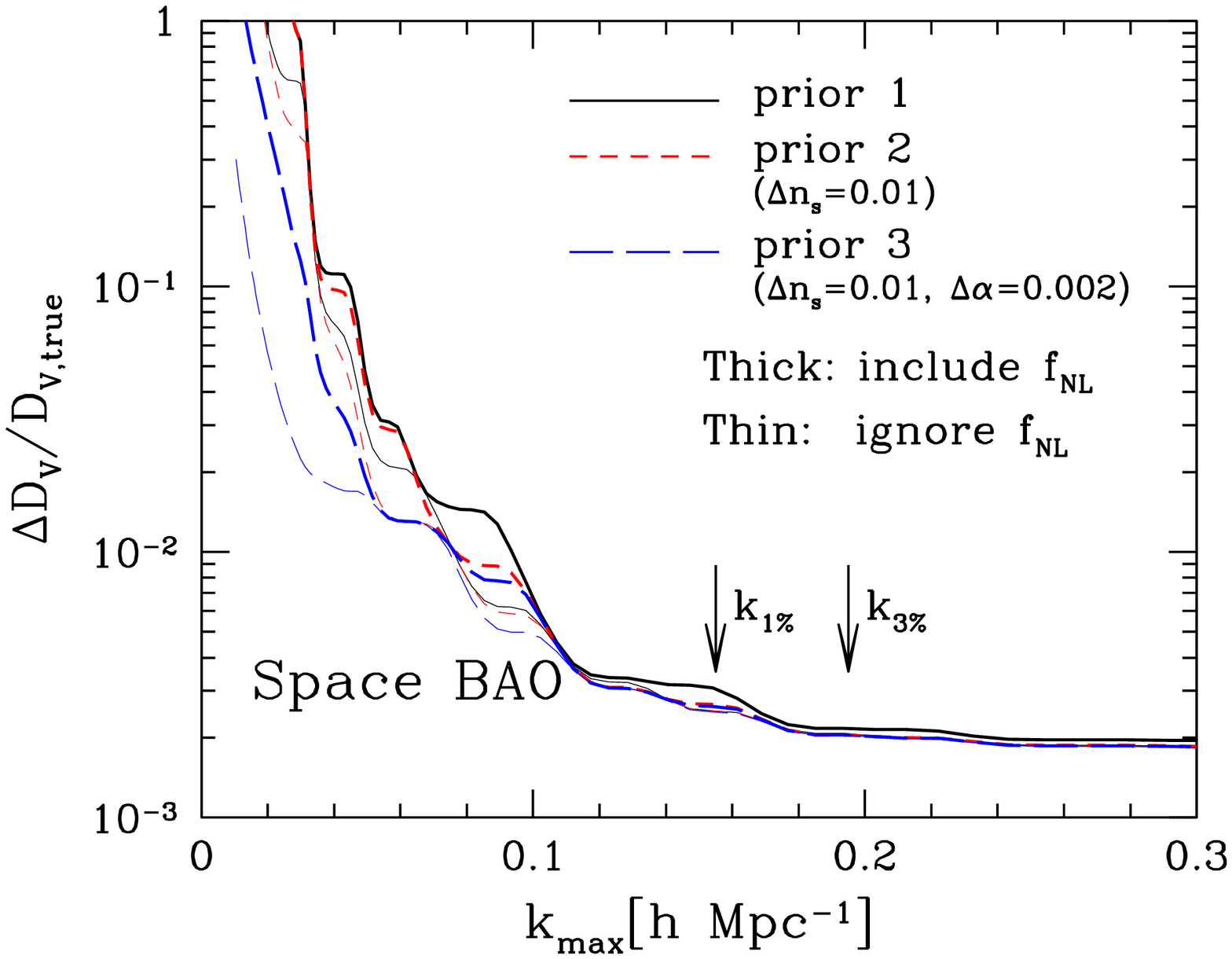}
\end{center}

\vspace*{-0.3cm}

\caption{Predicted 1-$\sigma$ (68\%C.L.) errors on the 
spectral index $n_s$ (left), running of the index $\alpha$ (middle), and 
distance scale $D_V/D_{V,{\rm true}}$ (right) as function of 
maximum wavenumber $k_{\rm max}$, assuming the survey parameters of  
$z=1.5$, $V_s=100h^{-3}$Gpc$^3$, $b_1=3.25$, and $n_{\rm gal}=10^{-4}
h^{3}$Gpc$^{-3}$, as an illustrative example of space-based BAO missions. 
Here, we specifically treat the local model of primordial non-Gaussianity. 
The solid (prior 1), short-dashed (prior 2), and long-dashed (prior 3) 
lines represent 
the results under the different priors (see text for details). 
Thick lines show
the one-dimensional errors marginalized over the four parameters 
(i.e., $n_s, \, \alpha, \, D_V/D_{V,{\rm true}},\, \fnl$), and 
thin lines represents the error excluding the non-Gaussian parameter, $\fnl$.  
\label{fig:1Derror_ns_nrun_Dv_ADEPT}}
\begin{center}
\includegraphics[height=5.1cm,angle=0]{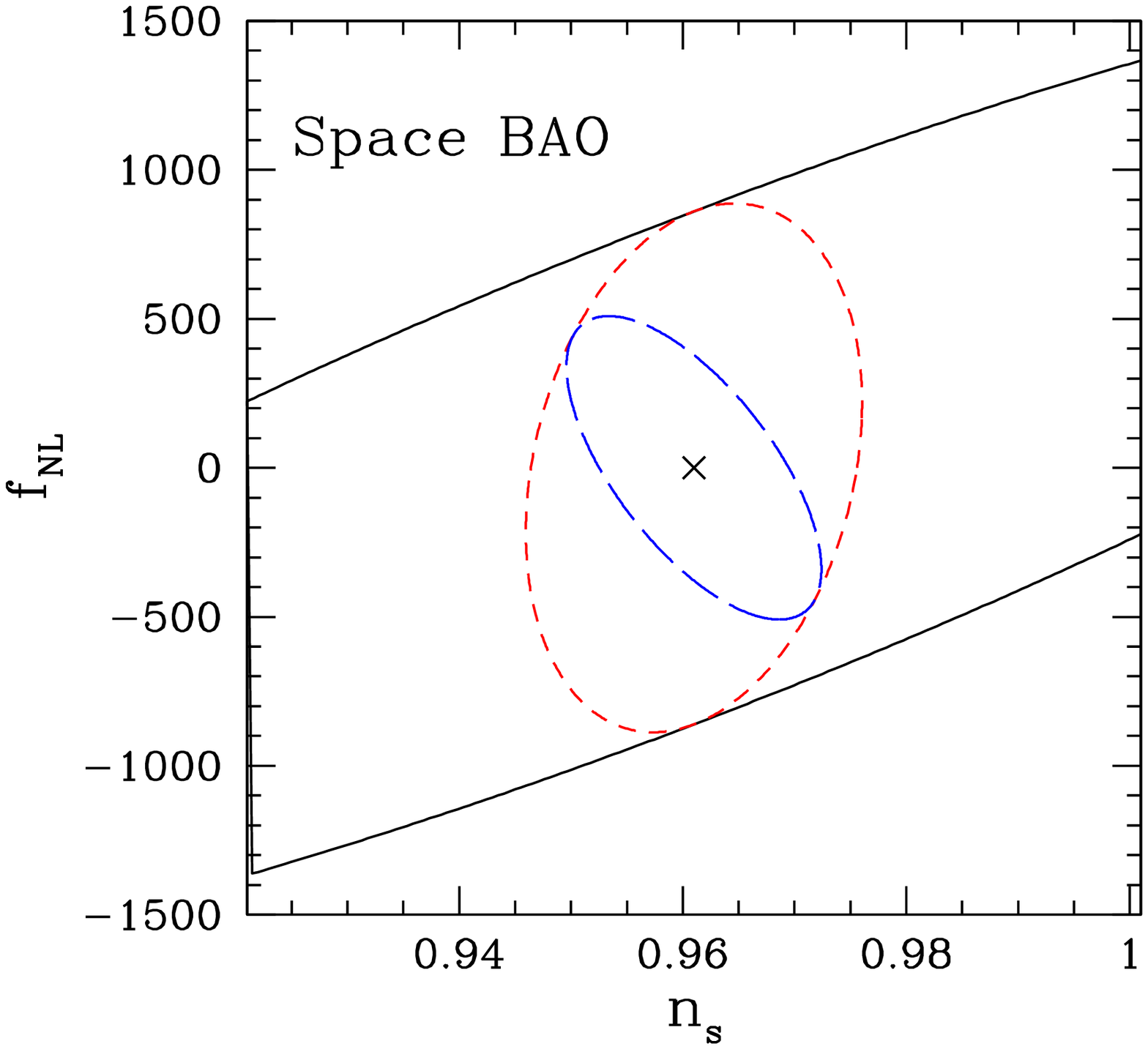}
\hspace*{0.1cm}
\includegraphics[height=5.1cm,angle=0]{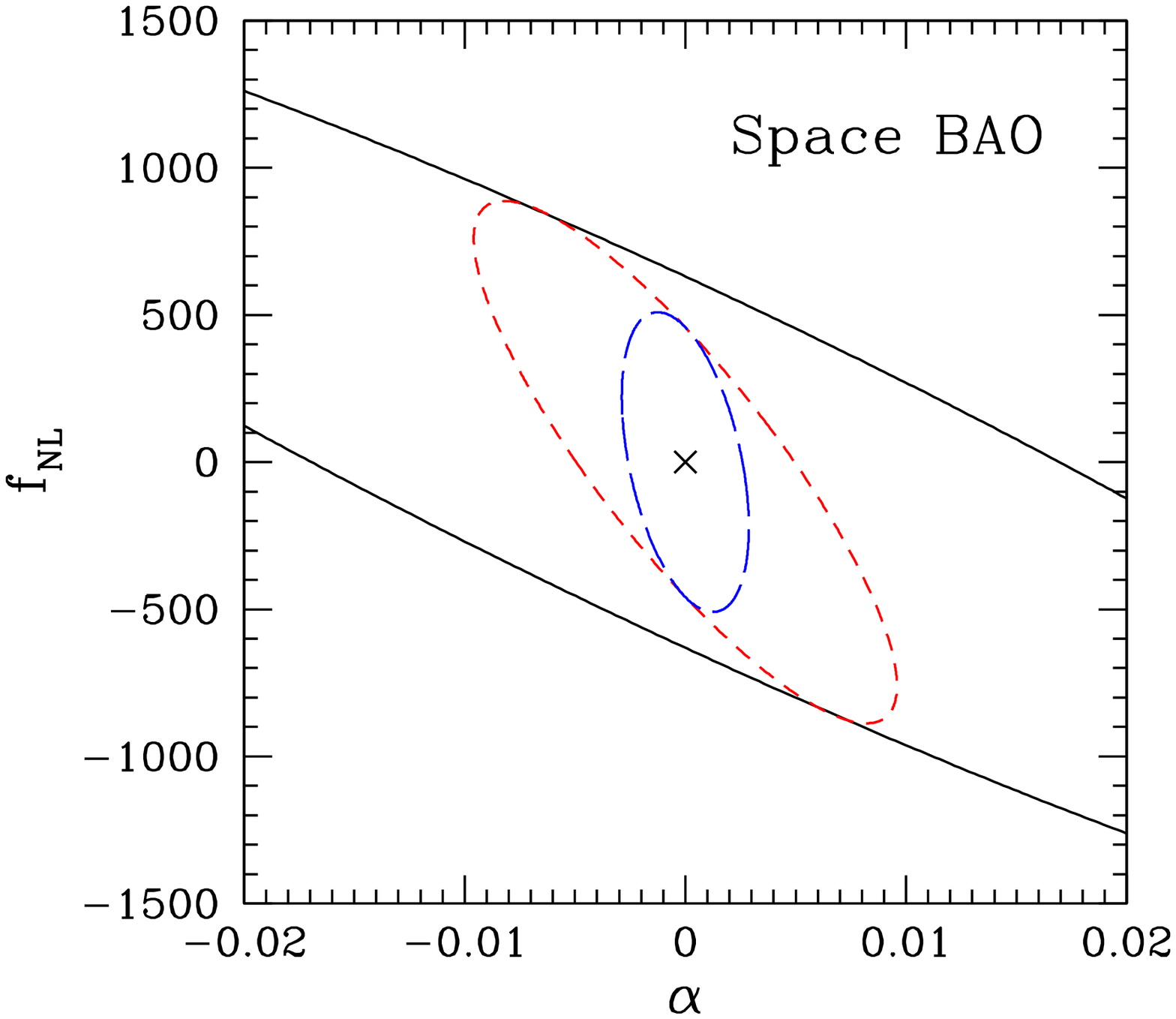}
\hspace*{0.1cm}
\includegraphics[height=5.1cm,angle=0]{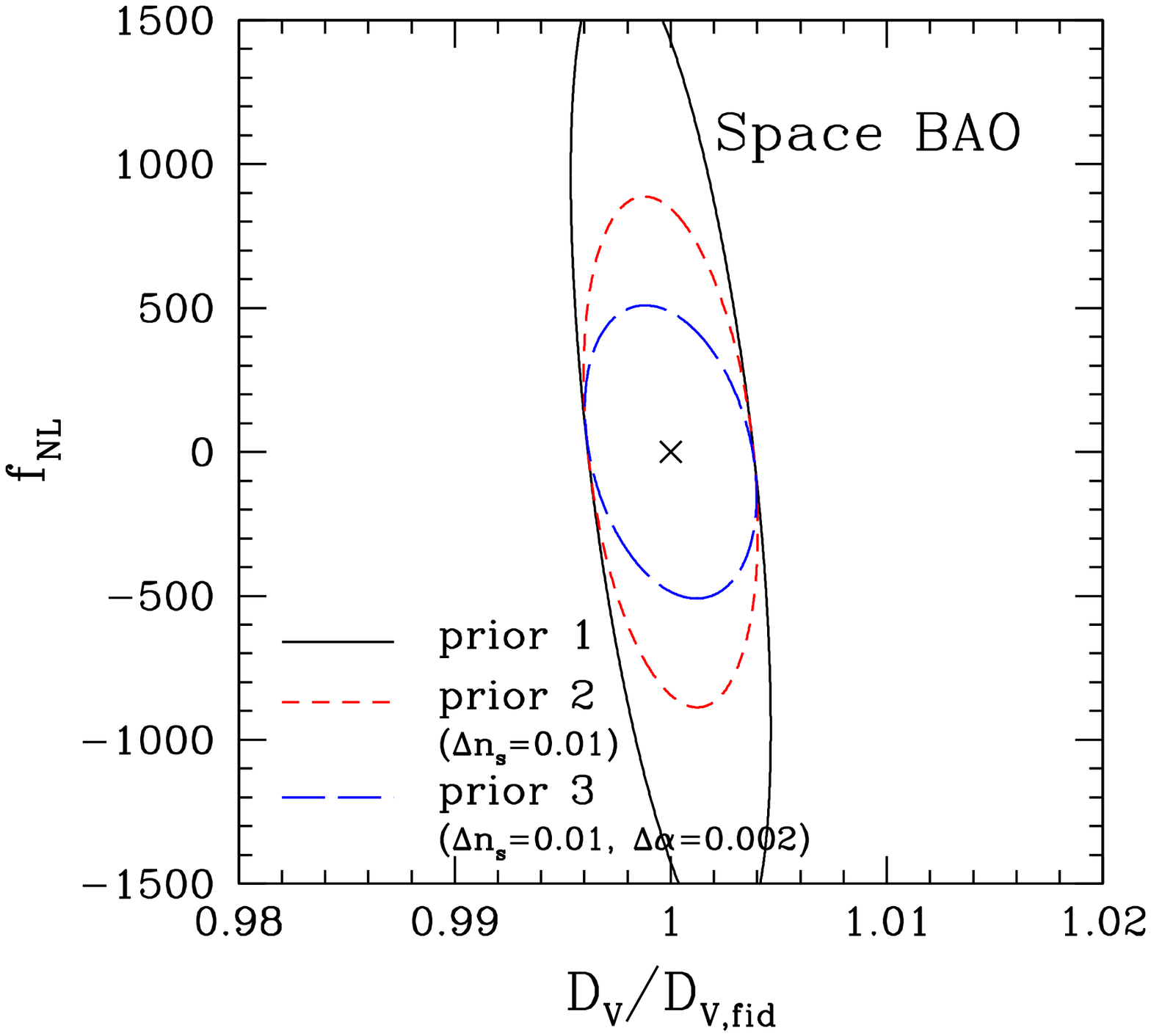}
\end{center}

\vspace*{-0.3cm}

\caption{Two-dimensional joint 68\%C.L. constraints on $\fnl$ and 
$n_s$ (left), $\alpha$ (middle), and $D_V/D_{V,{\rm true}}$ (right), 
assuming the local model of primordial non-Gaussianity. In deriving the 
constraints, we fix the maximum wavenumber 
to $k_{\rm max}=k_{1\%}\simeq0.155h$Mpc$^{-1}$ and adopt the survey 
parameters of a space-based BAO mission. The different lines indicate 
the results from imposing different prior information on $n_s$ and $\alpha$: 
no priors (solid): $\Delta n_s=0.01$ (short-dashed): 
$\Delta n_s=0.01$, $\Delta \alpha=0.002$ (long-dashed). 
\label{fig:2Derror_ns_nrun_Dv_ADEPT}}
\end{figure}

\begin{table}[b]
\caption{\label{tab:fisher_results}Marginalized one-dimensional errors 
(68\%C.L.) on $n_s$, $\alpha$, $D_V/D_{V,{\rm true}}$ and non-Gaussian 
parameters $\fnl$ and $\kappa$ from Fisher-matrix analysis adopting the 
survey parameters of a space-based BAO experiment}
\begin{ruledtabular}
\begin{tabular}{lccccc}
Fiducial model& $\Delta n_s$ & $\Delta\alpha$ 
& $\Delta(D_V/D_{V,{\rm true}})$ & $\Delta\fnl$ & $\Delta\kappa$$^\dagger$ \\
\hline
Local ($\fnl=0$) & 0.0075 & 0.0019 & 0.0026 & 335 & ---\\
Local ($\fnl=+100$) & 0.0075 & 0.0019 & 0.0026 & 338 & ---\\
Equilateral ($\fnl=+250,~\kappa=-0.3$) & 0.0072 & 0.0019 & 0.0025 & 596
& 0.48~(0.37)\\
Equilateral ($\fnl=+250,~\kappa=0$) & 0.0086 & 0.0020 & 0.0025 & 1306
& 0.48~(0.45)\\
Equilateral ($\fnl=+250,~\kappa=+0.3$) & 0.0082 & 0.0018 & 0.0025 & 2774
& 0.49~(0.49)\\
Equilateral ($\fnl=-150,~\kappa=-0.3$) & 0.0071 & 0.0019 & 0.0025 & 498
& 0.49~(0.43)\\
Equilateral ($\fnl=-150,~\kappa=0$) & 0.0085 & 0.0020 & 0.0025 & 1230
& 0.49~(0.48)\\
Equilateral ($\fnl=-150,~\kappa=+0.3$) & 0.0082 & 0.0018 & 0.0025 & 2753
& 0.50~(0.50)\\
\end{tabular}
\leftline{$^\dagger$ parentheses represent the marginalized 1-$\sigma$ 
error when further adopting the Gaussian prior of $\Delta\fnl=75$}
\end{ruledtabular}
\end{table}

Figure \ref{fig:1Derror_ns_nrun_Dv_ADEPT} shows the marginalized 1-$\sigma$ 
(68\%C.L.) errors on the spectral index $n_s$ (left), the running of the index 
$\alpha$ (middle), and the distance scale $D_V/D_{V,{\rm true}}$ (right) 
as function of cutoff scale, $k_{\rm max}$.  
Here, the solid lines represent the results 
assuming no prior information to these parameters ({\it prior 1}, solid). 
The short- and long-dashed lines correspond to the 
results taking account of the priors on $n_s$ and $\alpha$, 
which are expected to come from the upcoming Planck CMB experiment 
: $\Delta n_s=0.01$ ({\it prior 2}, short-dashed); $\Delta n_s=0.01$ 
and $\Delta\alpha=0.002$ ({\it prior 3}, long-dashed). We assume Gaussian 
priors for these cases.

Figure \ref{fig:1Derror_ns_nrun_Dv_ADEPT} indicates that 
the CMB priors play an important role for the determination of the 
primordial indices and distance scales. This has been repeatedly 
pointed out in the literature, but we emphasize that it is 
indeed crucial when we include the non-Gaussian parameter $\fnl$ 
in the analysis of cosmological parameter estimation. 
To clarify the influences, we exclude $\fnl$ from the Fisher-matrix 
analysis and evaluate the marginalized errors again. The results are 
plotted in thin lines. Comparing thick with thin lines, we see that 
the inclusion of the extra parameter $\fnl$ significantly worsen the 
constraints on the primordial index and its running. 
With strong priors on $n_s$ and 
$\alpha$ ({\it prior 3}), tighter constraints comparable to 
those ignoring primordial non-Gaussianity would be obtained. 
By contrast, the determination of distance scale $D_V/D_{V,{\rm true}}$ 
is insensitive to primordial non-Gaussianity if we choose 
the cutoff wavenumber $k_{\rm max}$ around $k_{1\%}$, labeled by the
vertical arrows. This is partly because the distance scale is mainly 
determined through the BAOs at linear scales, where the effect of primordial 
non-Gaussianity becomes negligible at $z\simeq1.5$.  
It is interesting to note that all the curves depicted in 
Figure \ref{fig:1Derror_ns_nrun_Dv_ADEPT} become almost constant
at $k\gtrsim k_{1\%}\simeq0.155h$Mpc$^{-1}$. Since this scale is slightly 
larger than the one inferred from the mean separation of galaxies 
($n_{\rm gal}^{1/3}\simeq0.07h$Mpc$^{-1}$), the constancy of 
the predicted errors implies that the  
shot-noise contribution becomes significant at $k\gtrsim k_{1\%}$.

In Figure \ref{fig:2Derror_ns_nrun_Dv_ADEPT}, we next plot the 
two-dimensional constraints on the non-Gaussian parameter $\fnl$ and one of 
the free parameters $n_s$, $\alpha$ and $D_V/D_{V,{\rm true}}$, marginalizing 
over the other remaining parameters. In this plot, 
the cutoff scale is fixed to 
 $k_{\rm max}=k_{1\%}$, and the 1-$\sigma$ (68\%C.L.) contours are 
shown for the three different priors. As expected from 
Figure \ref{fig:1Derror_ns_nrun_Dv_ADEPT}, 
the primordial index $n_s$ and its running $\alpha$ 
without CMB priors ({\it prior 1}) exhibit a strong 
degeneracy with the $\fnl$ parameter, while the degeneracy between $\fnl$ 
and distance 
scale is very weak. The uncertainty in the $\fnl$ parameter is therefore 
very large in this case. 
The errors can be reduced significantly when the strong 
constraints on $n_s$ and $\alpha$ are obtained ({\it prior 3}). 
However, the uncertainty in $\fnl$ is still $\Delta\fnl\sim500$, 
which is quite a bit larger than other observational techniques using 
the bispectrum or cluster abundance \cite{LMSV2007,SK2007,SSZ2004}. 
The uncertainty may be reduced if we increase the cutoff scale 
$k_{\rm max}$, although it is generally difficult to accurately predict 
the power spectrum at these scales, including the higher-order corrections. 
Hence, it is difficult 
to detect the non-Gaussian signals even from an idealistic 
large-volume survey.

Table \ref{tab:fisher_results} summarizes the results of the one-dimensional 
marginalized errors for various cases with different fiducial values, 
including the case of the equilateral model. 
Here, in addition to the priors on $n_s$ and $\alpha$ ({\it prior 3}),  
we further put a Gaussian prior on $\kappa$ with $\Delta\kappa=0.5$, 
the reason for which comes from the fact that $\kappa$ is the slow-roll 
parameter and should be restricted to $\kappa\ll1$. 
Because of the additional parameter, 
the resultant errors on $\fnl$ significantly increase for the equilateral 
model, especially with $\kappa>0$. By contrast, 
the errors on $n_s$, $\alpha$ and distance scale are hardly affected.

\begin{figure}[t]
\begin{center}
\includegraphics[height=6.5cm,angle=0]{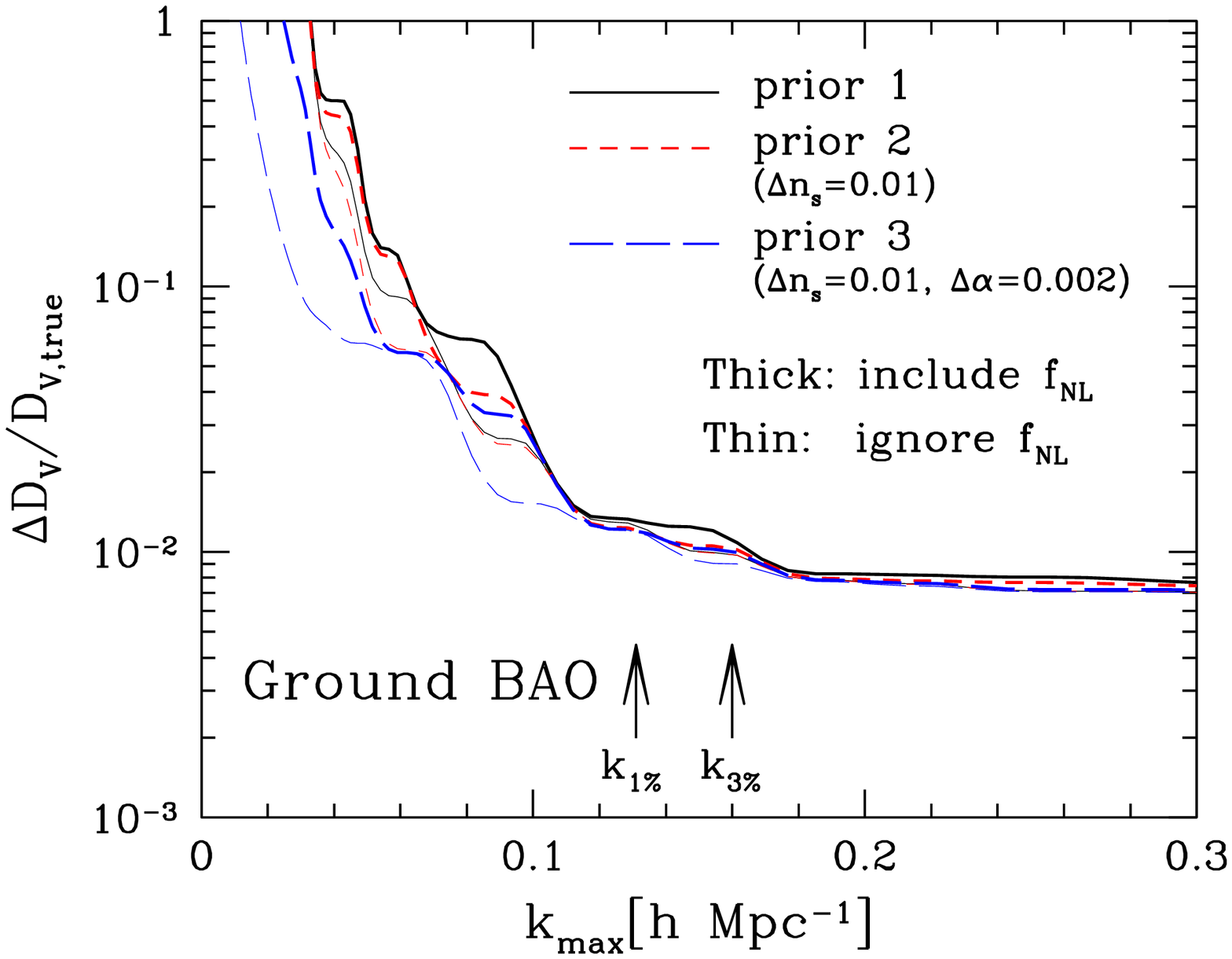}
\hspace*{0.5cm}
\includegraphics[height=6.5cm,angle=0]{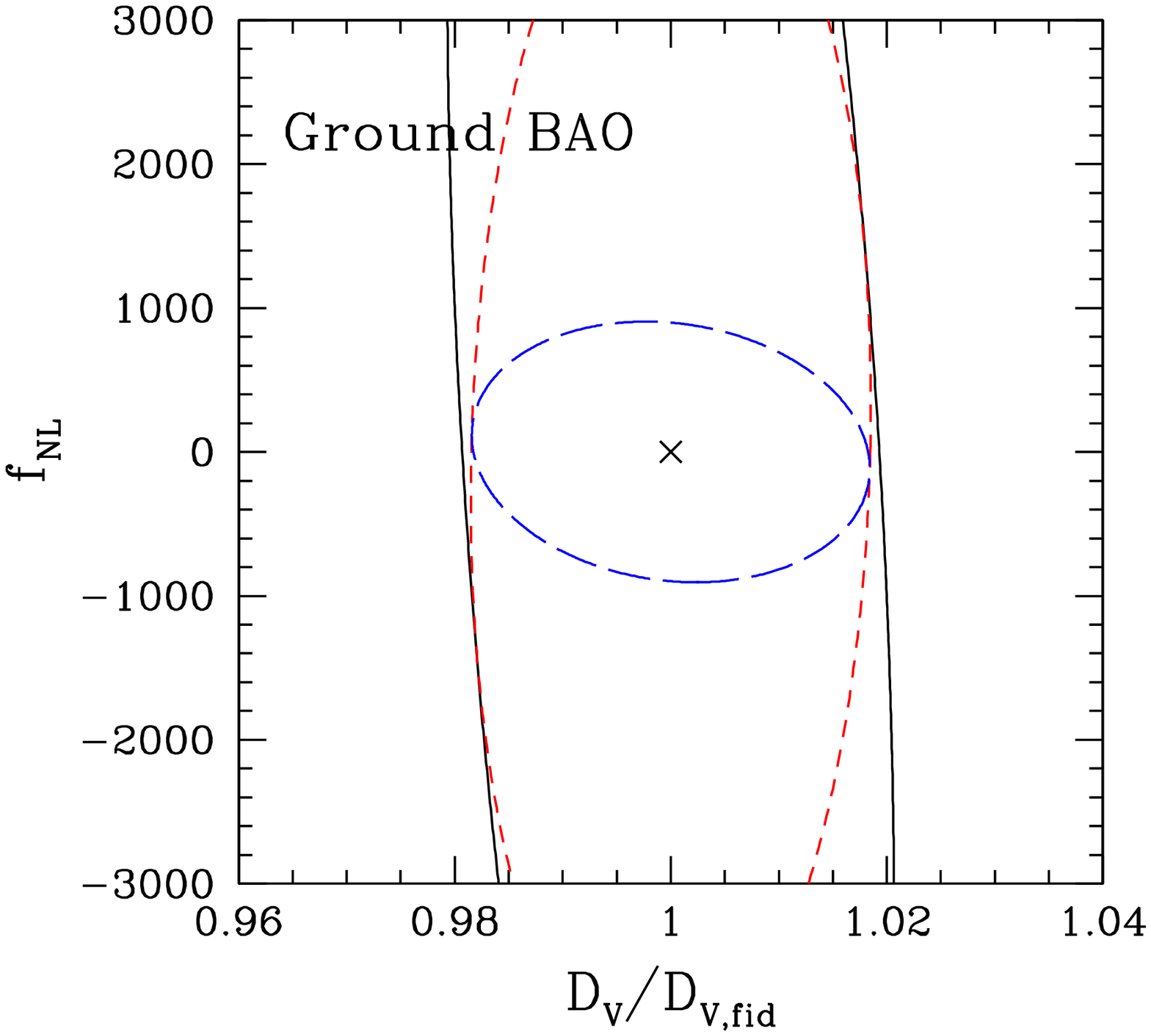}
\end{center}

\vspace*{-0.3cm}

\caption{Expected 1-$\sigma$ (68\%C.L.) error on the distance scale as 
a function of maximum wavenumber (left) and two-dimensional joint 
$68\%$ C.L. constraints on $\fnl$ of the local model and 
$D_V/D_{V,{\rm true}}$  
fixing the maximum wavenumber to $k_{\rm max}=k_{1\%}\simeq0.131h$Mpc$^{-1}$ 
(right). Here, we specifically adopt the 
survey parameters of 
$z=1$, $V_s=4$\,$h^{-3}$Gpc$^3$, $b_1=2.0$,  and $n_{\rm gal}=10^{-3}$ 
$h^{3}$Mpc$^{-3}$, as a representative example of ground-based BAO surveys. 
The meaning of the lines are the same as in 
Figures \ref{fig:1Derror_ns_nrun_Dv_ADEPT} and 
\ref{fig:2Derror_ns_nrun_Dv_ADEPT}.
\label{fig:errors_Dv_WFMOS}}
\end{figure}

Finally, as a representative example for a more near-term project, we 
consider the case of a ground-based BAO survey. 
The survey parameters we adopt are 
$z=1$, $V_s=4$\,($h^{-1}$Gpc)$^3$, $b_1=2$, and $n_{\rm gal}=10^{-3}$ 
($h^{-1}$Mpc)$^{-3}$ (\cite{DETF,WFMOS}, see also \cite{SK2007}). 
Figure \ref{fig:errors_Dv_WFMOS} shows the expected 
one-dimensional error on $D_V/D_{V,{\rm true}}$ (left) and the 
two-dimensional errors on $\fnl$ and $D_V/D_{V,{\rm true}}$ (right). 
Compared to the large volume of a space-based BAO experiment, 
the statistical errors of the measurement of the power 
spectrum are somewhat larger and thereby the precision of the 
distance-scale estimation is rather degraded. Nevertheless, 
the results are hardly affected by the primordial non-Gaussianity.

In conclusion, the presence of 
primordial non-Gaussianity potentially affects the estimation of 
the primordial index and the running, and the CMB-based priors on these 
parameters would be crucial, but the distance-scale measurement is relatively 
insensitive to it for the appropriate choice of $k_{\rm max}$. In the end, 
the primordial non-Gaussianity is hardly detectable from 
galaxy redshift surveys even with idealistically large survey-volume. 

\section{On the Local Biasing Prescription for Galaxy Power 
Spectrum}
\label{sec:galaxy_biasing}

So far, we have mainly dealt with the matter power spectrum and 
the galaxy power spectrum has been only considered under the 
simplified assumption of linear galaxy biasing. 
In reality, the statistical relation between galaxy and mass is 
generally complicated due to the non-linear nature of 
galaxy formation processes, and the linear biasing prescription 
would not hold even at large scales accessible to the future 
galaxy surveys.

Although the extension of the linear biasing prescription to include 
non-linear effects is rather non-trivial, a straightforward and frequently 
used prescription for galaxy biasing, coupled to perturbation theory, is 
the local biasing \cite{FG1993}, 
in which the galaxy density field $\deltag$ at a 
given position $\bfx$ is described as the local function of mass 
density field at the same position, 
i.e., $\deltag(\bfx)=f[\delta(\bfx)]$. On large scales of our interest, 
this can be expressed as the Taylor series expansion: 
\begin{eqnarray}
\deltag(\bfx|R)&=&f[\delta(\bfx|R)]=b_1\,\delta(\bfx|R) + 
\frac{b_2}{2}\,\bigl\{\,[\,\delta(\bfx|R)\,]^2 - 
\bigl\langle\,[\delta(\bfx|R)]^2\,\bigr\rangle\,\bigr\}
\nonumber\\
&&\quad\quad\quad\quad
 +\,\, \frac{b_3}{3!}\,\bigl\{\,[\,\delta(\bfx|R)\,]^3-
\bigl\langle\,[\delta(\bfx|R)]^3\,\bigr\rangle\,\bigr\}
\,+\,\cdots, 
\label{eq:local_bias}
\end{eqnarray}
where the quantities $\deltag(\bfx|R)$ and $\delta(\bfx|R)$ are the 
galaxy and mass density fields smoothed over the radius $R$ 
centered at the position $\bfx$. In the above expression, 
the coefficients $b_2$ and $b_3$ describe the non-linearity 
of galaxy biasing, which are incorporated into the 
galaxy power spectrum $P_{\rm gal}(k)$, through the perturbative 
calculation of the matter power spectrum 
(\cite{HVM1998,SSS2007,McDonald2006}, see also 
\cite{Taruya2000,Matsubara2008} for different parametrization schemes). 
Although the relation 
(\ref{eq:local_bias}) is just a phenomenological prescription, 
it has been recently applied to 
the characterization of the BAOs,  
and the model of galaxy power spectra 
has been tested against numerical simulations of 
halo/galaxy clustering in the case of Gaussian initial condition 
\cite{SSS2007,JK2008}.

When we consider primordial non-Gaussianity, 
several non-trivial corrections to the galaxy power spectrum $\Pg(k)$ 
appears, which can further alter the shape of the power spectrum on 
large scales. 
This has been first pointed out by McDonald~\cite{McDonald2008}. 
Here, we rephrase his finding and advocate its observational importance,  
together with potential problems.

Let us focus on the scales larger than the characteristic scale of BAOs, 
above which the higher-order perturbations can be safely neglected. 
Collecting the relevant terms in the expansion (\ref{eq:local_bias}), 
the power spectrum of the galaxy density fields smoothed over the 
radius $R$ may be expressed as 
\begin{eqnarray}
\Pg(k;z|R)= 
b_1^2\left\{\,D^2(z)P_0(k)+P^{(12)}(k;z)\,\right\}W^2(kR) + 
b_1b_2\,F(k;z,\fnl|R)+\cdots, 
\label{eq:P_gal_PT}
\end{eqnarray}
up to the third-order in mass density field, i.e., $\mathcal{O}(\delta_0^3)$. 
Here, $W(x)$ is the filter function defined in Fourier space. 
In equation (\ref{eq:P_gal_PT}), there are two types of non-Gaussian 
contributions. One is the term $P^{(12)}(k;z)$ coming from the matter 
power spectrum, and the other is the function $F(k;z,\fnl)$ arising from 
the non-linear mapping of the galaxy biasing. The explicit expression 
of the function $F$ is 
\begin{eqnarray}
&&F(k;z,\fnl|R)= D^3(z)\int\frac{d^3q}{(2\pi)^3}\,
M_\zeta(k)M_\zeta(q)M_\zeta(|\bfk-\bfq|)\,
\nonumber\\
&&\quad\quad\quad\quad\quad\quad\quad\quad\quad
\times\,\,W(kR)W(qR)W(|\bfk-\bfq|R)\,B_{\zeta}(-\bfk,\bfq,\bfk-\bfq).
\label{eq:def_F}
\end{eqnarray}
The functional form of $F$ is clearly different from the perturbative 
corrections of the matter power spectrum, $P^{(12)}$ (Eq.(\ref{eq:P12}) or 
(\ref{eq:P_ab^22(k)}) ). Hence, the presence of this 
term potentially leads to the scale-dependent galaxy biasing, relative 
to the matter power spectrum.

Before continuing, we note that the quantity $\Pg(k;z|R)$ manifestly 
depends on the smoothing radius $R$. In the usual sense, 
$\Pg(k;z|R)$ would not be directly related to the power spectrum derived 
from the unfiltered observations. It has been put forward 
in Ref.~\cite{SSS2007} 
that the unfiltered galaxy power spectrum $\Pg(k;z)$ can be 
recovered through the simple operation:
\begin{eqnarray}
\Pg(k;z) =\frac{\Pg(k;z|R)}{W^2(kR)}. 
\label{eq:def_Pg}
\end{eqnarray}
While the above definition still includes the smoothing radius, 
we regard equation (\ref{eq:def_Pg}) as a direct observable 
and evaluate it with the appropriate smoothing radius to see what happens. 
Below, we will separately treat the galaxy power spectra for the 
local and equilateral models of primordial non-Gaussianity.

\begin{figure}[h]
\begin{center}
\includegraphics[width=12cm,angle=0]{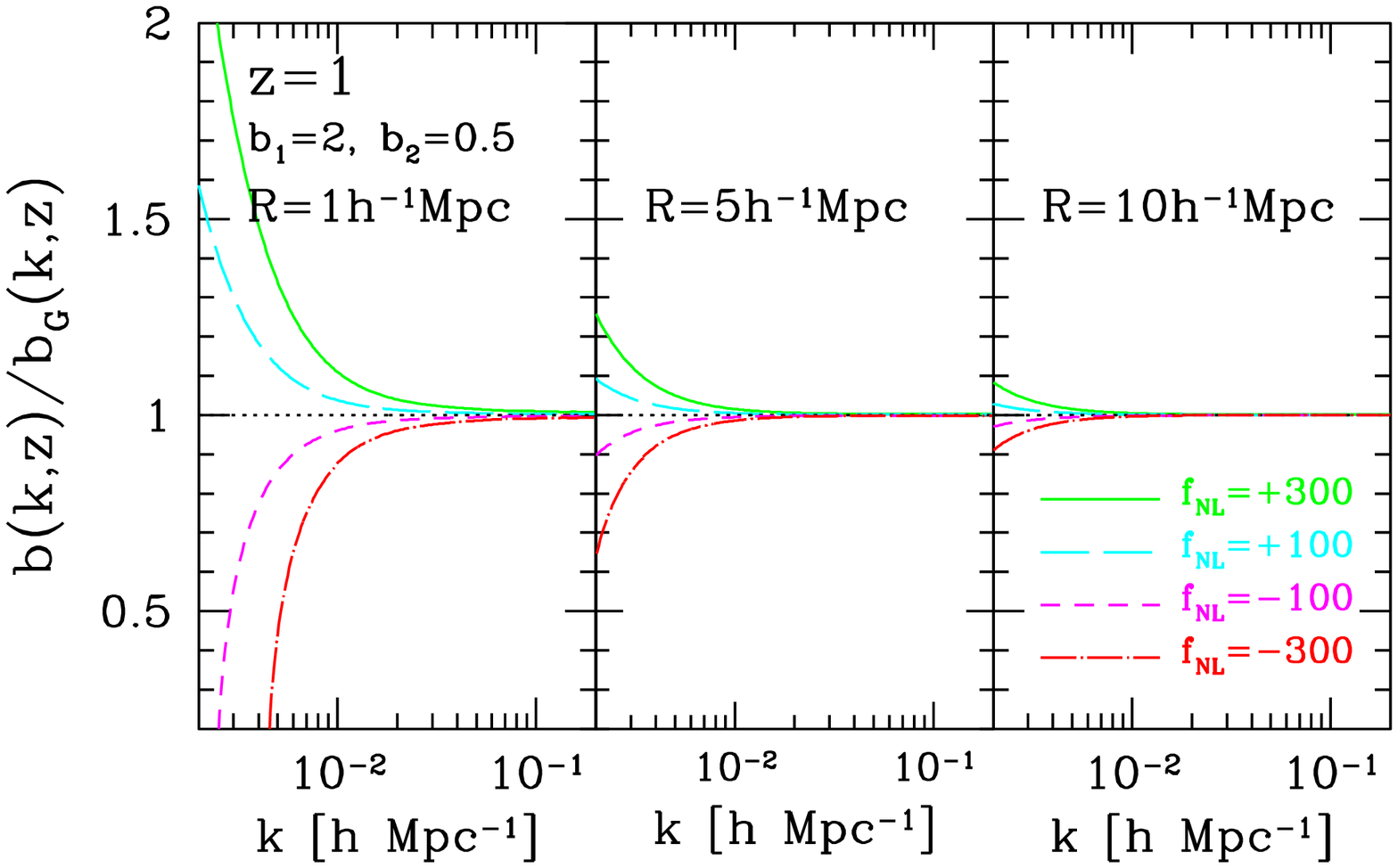}
\end{center}


\caption{Ratio of biasing factor, $b(k;z)/b_{\rm G}(k;z)$ 
  given at $z=1$, in the case of the local model. Gaussian smoothing is 
  adopted in order to 
  compute the biased power spectra. $R=1h^{-1}$Mpc (left);  
  $R=5h^{-1}$Mpc (middle);  $R=10h^{-1}$Mpc (right)
\label{fig:PT_bias_fNL_local}}
\begin{center}
\includegraphics[width=12cm,angle=0]{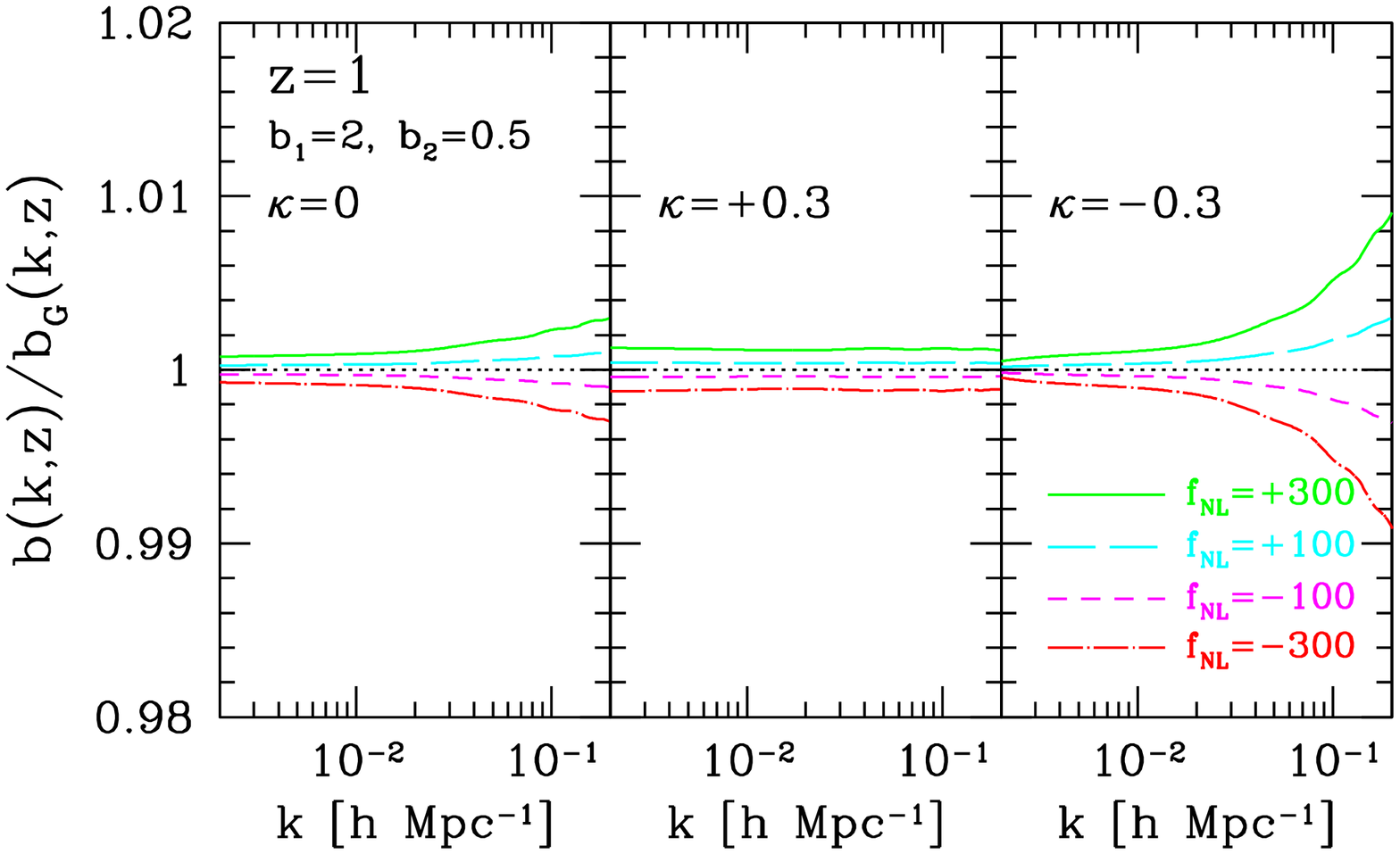}
\end{center}


\caption{Ratio of biasing factor, $b(k;z)/b_{\rm G}(k;z)$, given at $z=1$ 
  in the case of the equilateral model. In this plot, we do not consider 
  Gaussian smoothing. $\kappa=0$ (left); $\kappa=0.3$ (middle); 
$\kappa=-0.3$ (right). 
\label{fig:PT_bias_fNL_equilateral}}
\end{figure}

\subsection{Local model}

Figures~\ref{fig:PT_bias_fNL_local} 
plots the ratio of the biasing factor, $b(k;z)/\bG(k;z)$, plotted against 
the wavenumber, in the case of local model. 
Here, the biasing factor $b(k;z)$ is defined by 
\begin{eqnarray}
b(k;z)\equiv
\left\{\frac{P_{\rm gal}(k;z)}{P_{\rm mass}(k;z)}\right\}^{1/2}.
\label{eq:def_bias}
\end{eqnarray}
The function $\bG(k;z)$ is similarly defined as above, but  
with $\fnl=0$. For illustrative purposes, we specifically set the 
biasing coefficients to $b_1=2$ and $b_2=0.5$, and the results 
for $z=1$ are shown for different values of non-Gaussian parameter 
$\fnl$. Three different panels show the dependence of the smoothing radius, 
adopting 
Gaussian smoothing, $W(x)=\exp(-x^2/2)$, with radii 
$R=1h^{-1}$Mpc (left), $5h^{-1}$Mpc (middle), and $10h^{-1}$Mpc (right).

Clearly, the ratios of the biasing factor exhibit a strong 
scale-dependence, especially at $k\lesssim0.01h$Mpc$^{-1}$. 
Note that the function $F$ logarithmically diverges unless we 
introduce a large smoothing radius $R$. Accordingly, the strength of 
the scale-dependence sensitively depends on $R$ as well as $\fnl$. 
The origin of this scale-dependence can be deduced from a 
simple manipulation as follows. Substituting 
the bispectrum of the local model (\ref{eq:bispectrum_local}) 
into the expression 
(\ref{eq:P_gal_PT}), we take the limit $k\to0$. The correction 
$P^{(12)}(k)$ becomes negligibly small, and the term including 
the function $F$ is the only dominant contribution to the 
galaxy power spectrum. We have 
\begin{equation}
F(k;z,\fnl|R)\simeq  \frac{12}{5}\,\fnl\,D^3(z)\,P_0(k)\,
\frac{W(kR)\sigma^2(R)}{M_\zeta(k)};
\quad\quad \sigma^2(R)=\int\frac{d^3q}{(2\pi)^3}\,W^2(qR) P_0(q),
\end{equation}
where we used the fact that the function $M_\zeta(k)$ defined in 
equation (\ref{eq:M_zeta}) asymptotically behaves like 
$M_\zeta\propto k^2\,\, (k^0)$ in the limit of $k\to0$\, $(k\to\infty)$. 
Here, the quantity $\sigma(R)$ represents the rms amplitude of the linear 
density fluctuation $\delta_0$. Then, from (\ref{eq:def_Pg}) 
and (\ref{eq:def_bias}), we obtain
\begin{equation}
b(k;z)\simeq b_1 \left[ 
1+\frac{12}{5}\fnl\,\frac{b_2}{b_1} D(z)\frac{\sigma^2(R)}{M_\zeta(k)W(kR)}
\right]^{1/2}.  
\label{eq:bk_approx}
\end{equation}
The expression (\ref{eq:bk_approx}) contains a term inversely 
proportional to $M_\zeta(k)$, which leads to 
the strong scale-dependence at $k\to0$. Now, we recognize the fact 
that the scale-dependence of the biasing factor depends not only on 
$R$ and $\fnl$, but also on the ratio of biasing coefficients, 
$b_2/b_1$, and redshift, $z$.

It is interesting to note that similar kinds of behavior can be found 
in the halo biasing prescription \cite{DDHS2008,MV2008,SHSHP2008,AT2008}. 
While the halo biasing describes the 
clustering properties of 
virialized objects based on the Press-Schechter theory, 
the origin of the scale-dependence is qualitatively the same as that of 
the local biasing mentioned above. In this sense, the scale-dependent 
biasing may be a unique character of the local model of 
primordial non-Gaussianity and 
it would be an important observable of the large-scale structure.

Note, however, that the scale-dependent biasing factor in the local 
biasing seems problematic because of the 
logarithmic divergence at $R\to0$, which originates from the rms amplitude 
$\sigma(R)$ in the case of a CDM power spectrum. 
Further, the biasing factor $b(k;z)$ can be 
ill-defined for the negative $\fnl$ or $b_2$ and it eventually 
becomes complex at $k\to0$. McDonald \cite{McDonald2008} 
proposed a renormalization treatment to remove these difficulties by 
adding a non-local counter term to the relation (\ref{eq:local_bias}). 
Although the addition of a counter term might be a possible solution, 
uniqueness of this treatment seems is unclear, and thus the  
physical reason for the non-local counter term is unclear. 
In fact, as shown in next subsection, the scale-dependence of 
the biasing factor turns out to be 
model-dependent, and the logarithmic divergence disappears when we 
consider the equilateral model of primordial non-Gaussianity.

\subsection{Equilateral model}

In Figure~\ref{fig:PT_bias_fNL_equilateral}, 
we plot the ratio of the biasing factor, $b(k;z)/\bG(k;z)$, 
in the equilateral model. Here, instead of showing 
the dependence of smoothing radius, we neglect the effect of smoothing 
(i.e., $R\to0$) and just
examine the effect of scale-dependent non-Gaussianity by changing 
$\kappa$ to $0$ (left), $-0.3$ (middle) and $0.3$ (right).

In marked contrast to the local model, it turns out that  
the biasing factors $b(k;z)$ in the equilateral model are almost constant 
on large scales, and 
the linear deterministic biasing, $\Pg(k)=b_1^2P_{\rm mass}(k)$, is 
an excellent approximation to the galaxy power spectrum on large scales. 
These features are irrespective of the values of  $\kappa$, and  
can be deduced from the asymptotic behavior 
of the function $F$. In the $k\to0$ limit, we have 
\begin{eqnarray}
F(k;z,\fnl|R=0)&\simeq &\frac{18}{5}\,\fnl\,D^3(z)\,M_\zeta(k)\,
\nonumber\\
&&\quad\quad\times\,\,
\int\frac{d^3q}{(2\pi)^3}\,
\left(\frac{2q/3}{k_{\rm CMB}}\right)^{-2\kappa}\,\{M_\zeta(q)\}^2\,
\left[\,2\left\{P_\zeta(k)P_\zeta^5(q)\right\}^{1/3}-
\left\{P_\zeta(q)\right\}^2\,\right]. 
\end{eqnarray}
Recalling the power-law nature of the primordial spectrum 
$P_\zeta(q)\propto q^{n_s-4}$, the integral in the above equation 
converges if $n_s\sim1$ and $|\kappa|\lesssim 0.3$. 
As a result, the function $F$ scales as 
$F(k;z,\fnl|R=0)\propto M_\zeta(k)\propto k^2$ and 
the non-Gaussian contribution becomes negligibly smaller than 
the linear biasing term, $b_1^2P_{\rm mass}(k;z)$.

Therefore, as long as the equilateral model of primordial 
non-Gaussianity is concerned, it is hard to constrain $\fnl$ 
through the scale-dependent biasing factor. Although this 
conclusion comes from the assumption of 
the local biasing prescription, this would 
generally hold for other biasing schemes, including the halo 
biasing prescription. On the other hand, a big difference in 
the scale-dependent biasing factor between the 
local and equilateral models implies that the galaxy biasing is very 
sensitive to the shape of the primordial bispectrum. 
An essential reason for having a scale-dependent biasing only 
for the local-type primordial non-Gaussianity is that the amplitude of the 
primordial bispectrum with squeezed configuration, i.e., $B_{\zeta}(-\bfk,
\bfq, \bfk-\bfq)$ with $|\bfk| \ll |\bfq|$, is greater than that of
the equilateral case (see Eq.(\ref{eq:def_F})). 
The local-type primordial non-Gaussianity has 
fairly strong mode-correlations between small and large Fourier modes, 
and thereby the biasing, as a small-scale phenomenon, affects the power 
spectrum on very large scales. This generic feature might be 
very helpful to discriminate between the types of primordial 
non-Gaussianity. The detection of 
scale-dependent biasing may have interesting implication 
for the inflationary model, according to the single-field 
consistency relation \cite{CZ2004}. 
The consistency relation states that as long as the inflation dynamics 
is driven by a single scalar field, 
the primordial bispectrum in squeezed configuration becomes very 
small, irrespective of the type of non-Gaussianity. Thus, single-field 
inflation generically predicts scale-independent 
biasing on large scales. Therefore, 
if the presence of scale-dependent biasing on large scales is 
observationally verified, this strongly disproves the 
consistency relation, and the inflation dynamics may not be simply 
characterized by the single scalar field.

\section{Discussion and conclusions}

In this paper, we have comprehensively studied the effect of primordial 
non-Gaussianity on the power spectrum of large-scale structure. 
Using perturbation theory, we calculate the leading-order 
non-Gaussian corrections to the matter power spectrum through the 
non-linear mode coupling of the gravitational evolution. 
In the weakly non-linear regime, 
the signature of primordial non-Gaussianity in the matter power 
spectrum can be characterized by the primordial bispectrum. 
Adopting two representative models of the bispectrum (local and equilateral), 
we quantitatively estimate the non-Gaussian signals on the matter 
power spectrum. The primordial non-Gaussianity systematically 
enhances or suppresses 
the non-linear growth of the power spectrum amplitude at the $\lesssim$ 
1-2\% level.

We then explore the potential impact on the cosmological 
parameter estimations from future 
dedicated surveys for BAO measurement. Under 
the assumption of scale-independent linear biasing,
the Fisher-matrix analysis  reveals 
that while it is hard to detect the primordial non-Gaussianity to the 
level of current constraints, the inclusion of the non-Gaussian 
parameter $\fnl$ significantly degrades the constraints on 
the primordial spectral index $n_s$ and the running of the 
index $\alpha$. In this respect, the CMB prior information for 
$n_s$ and $\alpha$ as well as the non-Gaussian parameter 
is very crucial.  
On the other hand, determination of the distance scale $D_A(z)$ is 
rather insensitive to the presence or absence of primordial 
non-Gaussianity, and thus the characteristic scale of BAOs 
is a robust standard ruler.

We have also considered the effects of primordial non-Gaussianity on
the galaxy biasing. In the framework of local galaxy biasing, in which
the number density of galaxies is described by a local function of
the mass density field, we found that the non-linear mapping of galaxy
biasing can modulate the power spectrum and this sensitively depends
on the shape of non-Gaussianity. In the local model of primordial
non-Gaussianity, mode-correlations between large scales and small
  scales are fairly strong, and the non-Gaussian correction induces
the strong scale-dependent biasing on large-scales, while the
scale-independent linear biasing is preserved to a good accuracy in the
equilateral model. These remarkable properties may be very helpful in 
discriminating between the types of primordial non-Gaussianity, especially 
in connection with the single-field consistency relation. However,
the biasing factor in the local model exhibits an apparent divergence,
suggesting that the local biasing prescription may be incompatible
with the local model of primordial non-Gaussianity. Indeed, the 
local biasing
scheme is just a phenomenological parametrization and the validity of
this prescription itself has not yet been tested enough in the
presence of non-Gaussianity. Further study using simulations is
necessary for quantitative prediction of the non-Gaussian signature on
galaxy biasing.

\begin{acknowledgments}
We would like to thank Patrick McDonald and Shun Saito 
for comments and discussion, and Erik Reese for a careful reading of the 
manuscript. AT is supported by a Grant-in-Aid for Scientific 
Research from the Japan Society for the Promotion of Science (JSPS) 
(No.~18740132). TM acknowledges 
support from the Ministry of Education, Culture, Sports, Science, and 
Technology, Grant-in-Aid for Scientific Research (C) (No.~18540260, 2006). 
KK is supported by STFC and RCUK. This work was supported in part by 
Grant-in-Aid for Scientific Research on Priority Areas No.~467 
``Probing the Dark Energy through an Extremely Wide and Deep Survey with 
Subaru Telescope'', and JSPS Core-to-Core Program ``International 
Research Network for Dark Energy''.
\end{acknowledgments} 



\end{document}